\documentclass[11pt,a4paper]{article}
\usepackage{jcappub}
\pdfoutput=1
\usepackage{graphicx}
\usepackage{longtable}
\usepackage{amsmath}
\usepackage{amssymb}
\usepackage{hyperref}
\usepackage{dcolumn}
\usepackage{cleveref}
\usepackage{bm}
\usepackage{subfigure}
\usepackage{stackengine}
\usepackage{color}

\newcommand{\mpl}{{M_{\rm {pl}}}}

\newcommand{\dd}{\, {\rm d}}
\newcommand{\gsim}{\;\mbox{\raisebox{-0.5ex}{$\stackrel{>}{\scriptstyle{\sim}}$}
}\;}
\newcommand{\lsim}{\;\mbox{\raisebox{-0.5ex}{$\stackrel{<}{\scriptstyle{\sim}}$}
}\;}
\newcommand{\rs}{r_{\rm s}}

\newcommand{\pn}{\Phi_{\rm N}}

\newcommand{\oo}{\mathcal{O}}

\newcommand*\colvec[3][]{\begin{pmatrix}\ifx\relax#1\relax\else#1\\\fi#2\\#3\end{pmatrix}}
\newcommand{\tg}{\tilde{g}}

\newcommand{\nm}{{\mu\nu}}
\newcommand{\eff}{_{\rm eff}}

\newcommand{\mmm}{{_{\rm m}}}

\newcommand{\pmi}{\phi_{\rm min}}
\def\eea{\end{eqnarray}}
\def\bea{\begin{eqnarray}}
\allowdisplaybreaks
\usepackage{enumitem}
\setenumerate[1]{label=(\arabic*)}

\begin{document}
\title{A Compendium of Chameleon Constraints}
\author[a]{Clare Burrage}
\emailAdd{clare.burrage@nottingham.ac.uk}
\affiliation[a]{School of Physics and Astronomy, University of Nottingham, Nottingham, NG7 2RD, UK}
\author[b,c]{Jeremy Sakstein}
\emailAdd{jeremy.sakstein@port.ac.uk}
\affiliation[b]{Center for Particle Cosmology, Department of Physics and Astronomy, University of Pennsylvania 209 S. 33rd St., Philadelphia, PA 19104, USA}
\affiliation[c]{Institute of Cosmology and Gravitation,
 University of Portsmouth, Portsmouth, PO1 3FX, UK}

\abstract{
The chameleon model is a scalar field theory with a screening mechanism that explains how a cosmologically relevant light scalar can avoid the constraints of intra-solar-system searches for fifth-forces. The chameleon is a popular dark energy candidate and also arises in $f(R)$ theories of gravity. Whilst the chameleon is designed to avoid historical searches for fifth-forces it is not unobservable and much effort has gone into identifying the best observables and experiments to detect it. These results are not always presented for the same models or in the same language, a particular problem when comparing astrophysical and laboratory searches making it difficult to understand what regions of parameter space remain. Here we present combined constraints on the chameleon model from astrophysical and laboratory searches for the first time and identify the remaining windows of parameter space. We discuss the implications for cosmological chameleon searches and future small-scale probes. 
}
\maketitle

\section{Introduction}

The expansion of the universe is currently accelerating. The parsimonious explanation for this, that a cosmological constant should be included in 
Einstein's equations combined with quantum mechanical calculation of its value, spectacularly fails to predict a universe in which we could live. 
Predictions for the value of the cosmological constant are typically so large that galaxies would not even be able to form. No conclusive solution 
to the cosmological constant problem exists but a common side effect of current attempts at a solution is the introduction of new light scalar 
degrees of freedom. These light scalars in turn cause their own problems as they will mediate a new long-range fifth-force which is not observed to a 
high degree of precision in solar system or laboratory searches. As the cosmological constant problem is the result of the vacuum fluctuations of 
standard model particles, it is expected that couplings between such a scalar and matter are unavoidable.

These scalars are known as dark energy, and their existence can be reconciled with the lack of observation of a fifth-force through the use of 
screening mechanisms which allow the scalar to vary its properties with the local environment. The archetypal example of this is the chameleon 
mechanism, which allows the mass of the scalar to increase in dense environments (this is analogous to how the Debye mass of a photon in a plasma 
increases with the charge density). The low density of intergalactic space means that the chameleon can have a light mass and mediates a long-range 
force on cosmological scales but has a high mass inside the test masses used to search for fifth-forces, suppressing its effects there. 

A wide range of experiments are either currently searching for the chameleon, or will do in the near future; ranging from table top laboratory 
experiments using cold atoms to satellite missions aiming to map the formation and evolution of structure in our universe. These experiments all 
exploit the fact that the chameleon varies its mass depending on the environment so a suitably designed experiment or tailored astrophysical 
observable has the potential to be able to detect its presence. To date, the various communities studying the chameleon have 
used a variety of different notations meaning that constraints are not easy to compare and we are currently unable to identify the best 
strategy for detecting or excluding the most popular current model of dark energy. Here, we address this lack by providing combined constraints on 
the chameleon from both laboratory measurements and astrophysics. As a result, we are able to identify the remaining areas of parameter space and the 
experiments with the best prospects for covering the remaining space. The results we combine here have all previously been published. The novel contribution of this article is the combination of constraints that we present in Figure \ref{fig:cons}, and the extension of some existing constraints to a broader range of chameleon models. We do not report every experiment that has ever placed constraints on the chameleon model, focusing instead on the current state of the field so that we can present the most up to date exclusion plots for the chameleon. We hope that this will prove useful to future attempts to search for the chameleon and allow such experiments to be targeted at the most viable models. 

\section{Chameleon Dark Energy}\label{sec:CDE}

The chameleon is a model of a canonical scalar field sourced by both matter and a non-trivial scalar potential \cite{Khoury:2003aq,Khoury:2003rn}. We will start 
from the equation of motion for the chameleon scalar $\phi$
\begin{equation}\label{eq:chameomrel}
\Box\phi= \frac{d V(\phi)}{d\phi} + \frac{\rho}{M}, 
\end{equation}
where $\rho$ is the density of non-relativistic matter and $M$ is a new mass scale characterising the strength of the coupling to matter. In appendix \ref{sec:covtheory} we provide a complete 
covariant theory for the theoretically inclined reader and derive some useful results but all we will require for our purposes is the scalar force 
\begin{equation}\label{eq:force}
 \vec{F}_\phi=\frac{\vec{\nabla}\phi}{M}.
\end{equation}

We start by considering what happens if the scalar is not a chameleon and has no ability to vary its mass. If we were to choose a simple mass term 
for the potential $V(\phi)= (1/2)m_0^2 \phi^2$ we would have a standard Yukawa scalar and the ratio of the scalar to gravitational forces around a 
massive spherical test mass is
\begin{equation}
 \frac{F_5}{F_N}=2\left(\frac{\mpl^2}{M}\right)e^{-m_0R}.
\end{equation}
Yukawa modifications of the inverse-square law are tightly constrained by laboratory fifth-force searches and Lunar Laser Ranging (see e.g. 
\cite{Adelberger:2003zx,Adelberger:2005vu,Kapner:2006si,Murphy:2012rea}) and one typically needs to tune 
$m_0\gg H_0$, in which case the scalar has little to nothing to say about dark energy, or to tune $M\gg \mpl$, in which case interactions between 
the scalar and matter would be extremely difficult to observe. Choosing $M\gg \mpl$ is also unsatisfactory because in order to trust a theory with such a high 
coupling constant it would be necessary to have knowledge of physics above the Planck scale. 

The scalar field in the analysis above is tightly constrained precisely 
because the equation of motion is linear: the mass of the field in the solar system is identical to the cosmological mass. 
Chameleon models circumvent this by choosing a non-linear potential \cite{Khoury:2003rn}. We can consider the dynamics of the scalar field as being governed by an effective potential
\begin{equation}
V_{\rm eff}(\phi)=V(\phi)+\frac{\phi \rho}{M}.
\end{equation}
Any choice of $V(\phi)$ that results in a minimum for the effective potential where the mass of small fluctuations about this minimum 
increases with $\rho$ is a valid choice for the chameleon potential. Specifically, this condition forbids us from choosing $V(\phi)\propto 
\phi^2$ or $ V(\phi)\propto \phi$. 
The most common choice is to assume that the potential has an inverse  power law form; specifically
\begin{equation}\label{eq:SP}
 V(\phi)=\tilde{\Lambda}^4+\frac{\Lambda^{4+n}}{\phi^n}.
\end{equation}
This potential has a density-dependent minimum given by
\begin{equation}\label{eq:pmi}
 \pmi(\rho)= \left(\frac{nM\Lambda^{4+n}}{\rho}\right)^{\frac{1}{n+1}}
\end{equation}
and the mass of fluctuations around this minimum is
\begin{equation}
 m\eff^2(\rho)=V_{{\rm eff}\, \phi\phi}=n(n+1)\Lambda^{4+n}\left[\frac{\rho}{nM\Lambda^{4+n}}\right]^{\frac{n+2}{n+1}}
\end{equation}
so the mass is an increasing function of the density provided that either $n>0$, $-1<n<0$ or 
$n$ is an even negative integer i.e. $n=-4,-6,-8,\ldots$. The case $n=0$ is simply a cosmological constant, the case $n=-1,-2$ does not allow the 
mass to vary with the density, and there is no minimum when $n=-3,-5,-7,\ldots$. 

\begin{figure*}[t,h]
\begin{center}
\stackunder{
\includegraphics[width=0.48\textwidth]{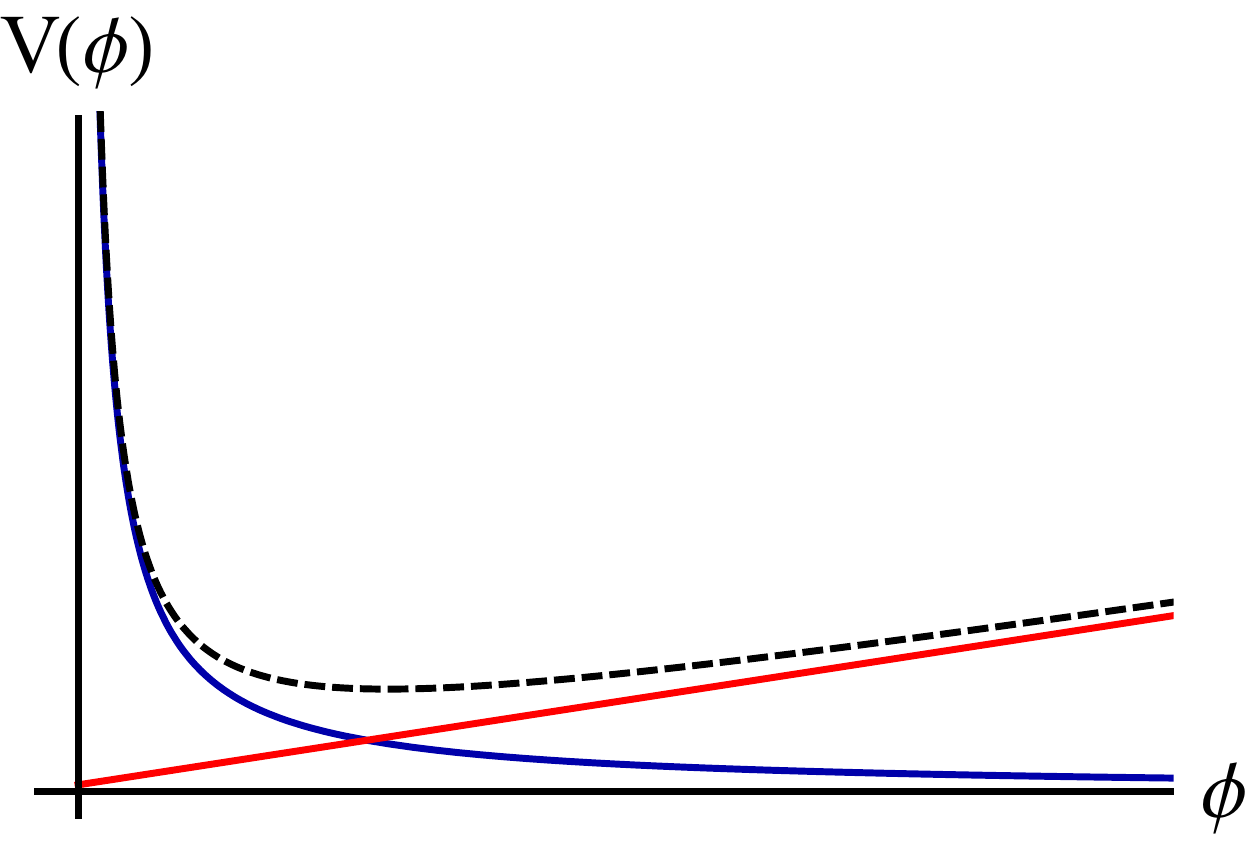}}{$n>0$, low density}
\stackunder{
\includegraphics[width=0.48\textwidth]{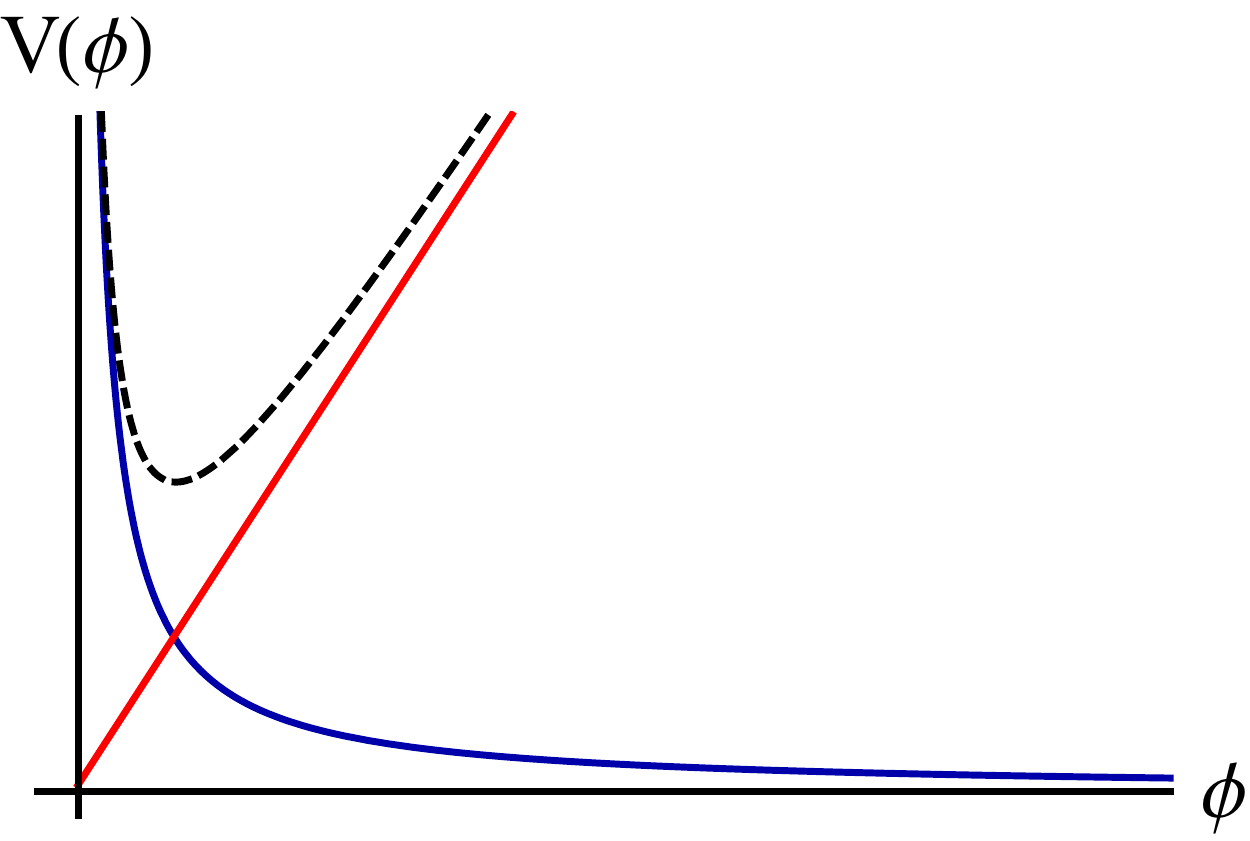}}{$n>0$, high density}
\stackunder{
\includegraphics[width=0.48\textwidth]{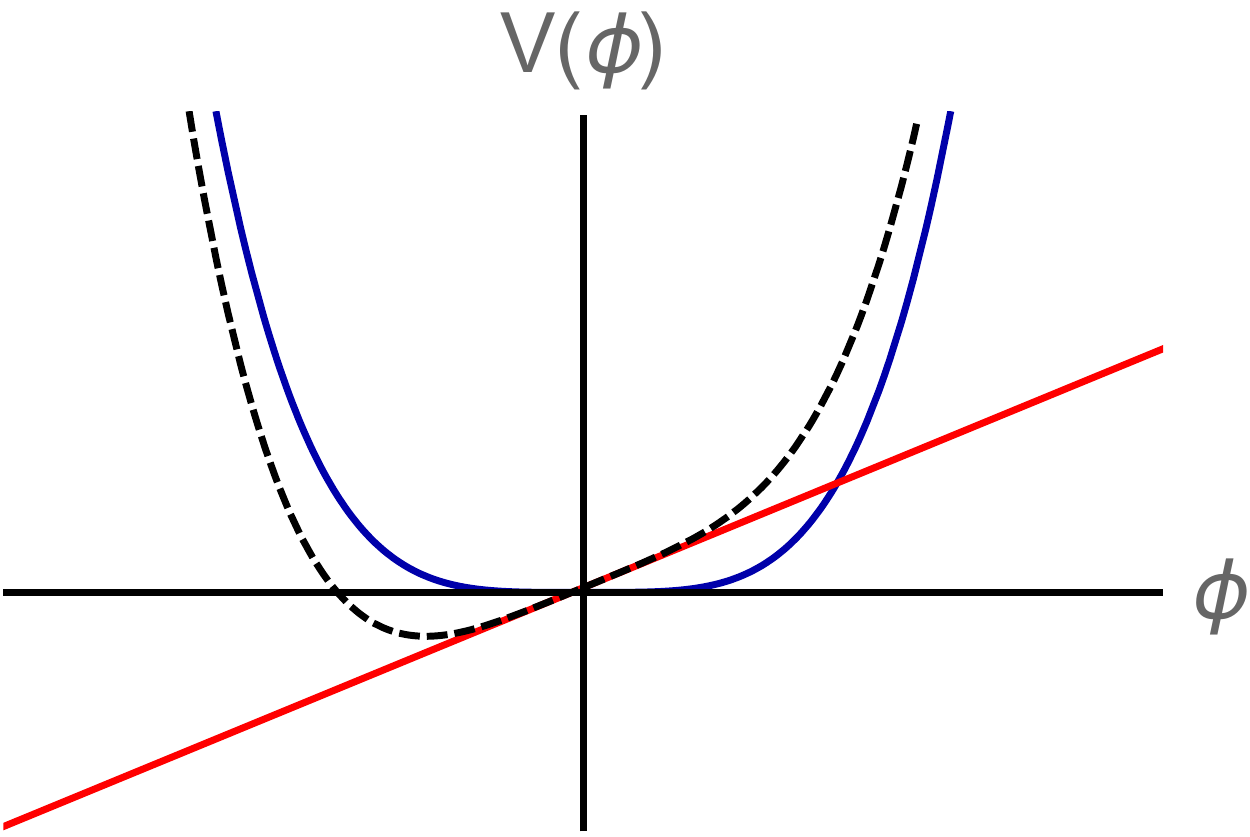}}{ $n<0$, low density}
\stackunder{
\includegraphics[width=0.48\textwidth]{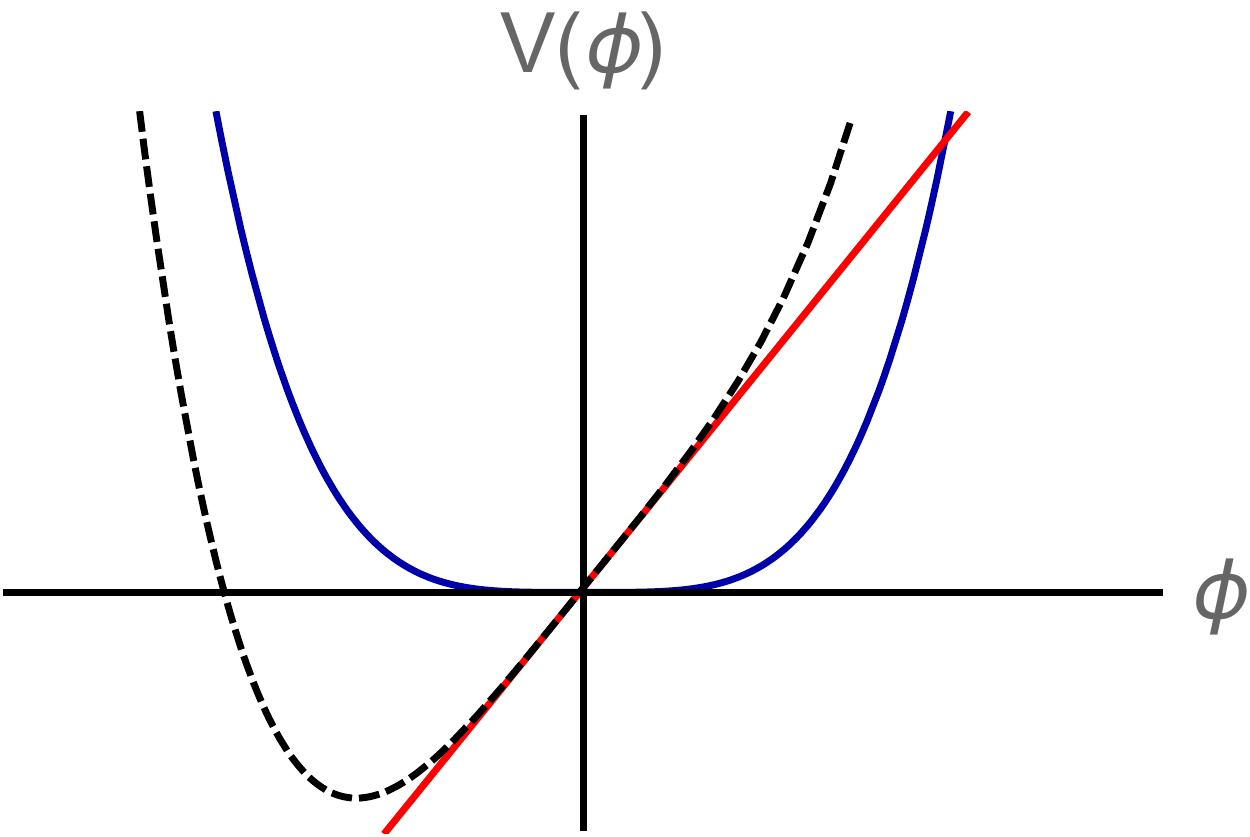}}{$n<0$, high density}
\caption{Sketch of the chameleon effective potential, for positive $n$ (upper panels) and negative $n$ (lower panels). Potentials are shown for low density environments (left panels) and high density environments (right panels). The blue line indicates the bare potential, the red line the 
contribution from the coupling to matter, and the black dashed line the sum of the two contributions.} \label{fig:potential}
\end{center}
\end{figure*}

The effective potential for positive and negative powers are shown in figure \ref{fig:potential} where one can clearly see that the curvature near the minimum is larger in high-density 
environments. The 
density-dependence allows for models where the scalar is light on cosmological scales so that it can drive dark energy but heavy in the solar system 
so that solar system tests are avoided. It is this blending in with the environment that has inspired the name \textit{chameleon}. 

\begin{figure}
\begin{center}
\includegraphics[width=0.45\textwidth]{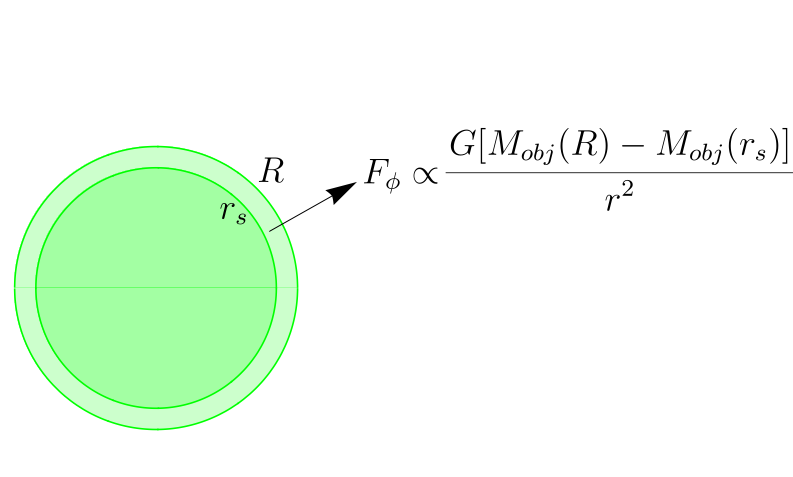}
\caption{The thin shell  mechanism. The non-linear potential \eqref{eq:SP} ensures that the mass inside the screening radius, $\rs$ does not source the field. The fifth-force is only sourced by the mass inside the light green thin shell indicated with by the arrow. In contrast, the Newtonian force is sourced by the entire mass of the object and so the fifth-force is highly suppressed in comparison.} \label{fig:sph}
\end{center}
\end{figure}

The varying mass allows the chameleon to avoid searches for fifth-forces through the so called ``thin-shell'' effect shown in figure \ref{fig:sph}. Consider a static, spherically symmetric source placed 
into some larger homogeneous background with density $\rho_0$ and field $\phi_0$, which could be the value that minimises the potential in the background, or could be set by the geometry of the experiment.  Deep inside the 
object, the field dynamics drive it towards the high-density minimum. Moving radially outward, the large mass ensures that the field will be frozen at 
this minimum. At large distances from the object, the field tends towards $\phi_0$ and so one expects that there is some radius, the screening radius 
$\rs$, at which the field gets pulled out of the minimum of the effective potential inside the source and starts to evolve towards the background 
value $\phi_0$. This results in the following form for the scalar field outside the object 
\cite{Khoury:2003rn,Hui:2009kc,Chang:2010xh,Davis:2011qf,Sakstein:2013pda,Sakstein:2015oqa}
\begin{equation}\label{eq:thinshell}
 \frac{\phi'}{M}\approx 2\left(\frac{\mpl}{M}\right)^2\frac{G\left(M_{\rm obj}-M_{\rm obj}(\rs)\right)}{r^2}e^{-m_0(r-R)},
\end{equation}
where $M(\rs)$ is the mass enclosed within the screening radius and $M_{\rm obj}$ is the object's total mass. We give an expression for the screening radius in the general case in equation (\ref{eq:screenrad}) and in the simplest  case, where the source has constant density in Equation (\ref{eq:rsconst}). 
The freezing of the field inside the object has the result that the exterior profile 
is sourced only by the mass outside the screening radius. When $\rs=0$ the field is sourced by the entire object and we are left with the situation 
described above where the theory describes a simple Yukawa interaction. This is the unscreened case that requires fine-tuning to satisfy solar system 
bounds. Conversely, when $\rs\approx R$ the field is sourced only by a very thin shell near the surface. The ratio of the fifth- to 
Newtonian-force is
\begin{equation}\label{eq:f5}
 \frac{F_5}{F_N}=2\left(\frac{\mpl}{M}\right)^2\left(1-\frac{M_{\rm obj}(\rs)}{M_{\rm obj}}\right)e^{-m_0(r-R)}
\end{equation}
and we see how screening naturally suppresses the chameleon fifth-force without the need to tune $M$ or $m_0$. The force profile highlights an 
important property of chameleon theories: they violate the equivalence principle.\footnote{This is an emergent property of solutions of the theory. There is no violation of the equivalence principle at the level of the Lagrangian. } Since the force depends on the mass enclosed within the 
screening radius, the acceleration of an extended object depends not only on its total mass but on its composition, which sets the value of $\rs$ 
(see section \ref{sec:astro} below). For this reason, objects with identical masses but different internal structures fall at different rates in the 
presence of external fields, signifying a breakdown of the equivalence principle (see \cite{Hui:2009kc} for an expanded discussion on this). 

One can 
see from equation \eqref{eq:f5} that whether or not an object screens depends on 
whether the field can reach the minimum inside an object, or, equivalently and perhaps more intuitively, the existence and location of the screening 
radius. We will refer to screening that occurs in this way as self-screening. In addition, due to the non-linear nature of the equations, the field profile for two nearby objects is not simply a 
superposition of the profiles for the individual bodies and this can lead to the environmental screening of smaller objects by larger ones. We refer to this as environmental screening, and it will have important implications for experimental tests of chameleons, which we will discuss further below.

Unfortunately, any model that successfully screens in the solar system cannot have a cosmological mass that is light enough to drive the cosmic 
expansion \cite{Wang:2012kj,Brax:2011aw}. Indeed, big bang nucleosynthesis constraints on the variation of standard model particle masses require the 
cosmological field to track its (time-dependent) minimum \cite{Brax:2004qh} and, at the level of background cosmology, the dominant term in equation 
\eqref{eq:SP} is the cosmological constant. For this reason, special attention is often paid to the particular choice $\Lambda = 2.3\times10^{-3}$ eV 
i.e. the dark energy scale. Indeed, many of the experiments discussed below make this choice from the outset. Whilst we focus our attention around this scale, we do not restrict ourselves to precisely this value, as the mechanism which solves the cosmological constant problem is as yet unknown, and until we know how the chameleon is related to such a solution we should allow for some variation between the cosmological constant scale and the scale controlling chameleon self interactions. In 
light of this, chameleon models should be viewed as alternatives to $\Lambda$CDM that predict an identical background expansion but exhibit novel 
effects and make differing predictions on smaller scales.

Two important specific cases are the case $n=1$ due to it being the simplest and most well-studied model, and $n=-4$, where  the self-interaction mass-scale 
$\Lambda$ is absent and the potential is given by
\begin{equation}\label{eq:p^4}
 V(\phi)=\lambda \phi^4,
\end{equation}
which is re-normalisable. In this case, $\lambda\sim\mathcal{O}(1)$ is seen as being natural since smaller values are fine-tuned and larger ones are 
strongly coupled. Regardless of the choice of bare potential, the full chameleon theory will never be renormalisable because the coupling 
between the scalar and matter fields necessarily introduces higher order operators. Therefore, the chameleon should be thought of only as a low energy 
effective theory valid below some cut-off and not as a fundamental description of the universe. Regions of parameter space where Coleman-Weinberg 
type quantum corrections to the chameleon mass could be kept under control in fifth-force experiments were identified in \cite{Upadhye:2012vh}. Lack 
of control over high energy quantum corrections to the theory has also been shown to mean that it is not currently possible to consistently evolve the 
chameleon theory through the radiation dominated epoch of the universe \cite{Erickcek:2013oma,Erickcek:2013dea}\footnote{Although see 
\cite{Padilla:2015wlv} for a potential UV extension that can evade quantum corrections.}.

The chameleon also arises in $f(R)$ theories of modified gravity where the Einstein-Hilbert action is replaced by one of the form
\begin{equation}
 S=\int\dd^4 x\sqrt{-g}\frac{R+f(R)}{16\pi G}.
\end{equation}
It can be shown that these are equivalent to a theory of standard Einstein Hilbert gravity plus a scalar field which couples to 
matter with strength $M=\sqrt{6}\mpl$ \cite{Brax:2008hh}. Therefore, the only viable models of $f(R)$ gravity must screen their fifth-forces using 
the chameleon mechanism 
\cite{Brax:2008hh}. The quintessential paradigm for $f(R)$ chameleons is the model of \cite{Hu:2007nk}
\begin{equation}
 f(R)=-a\frac{\mu^2}{1+(R/\mu^2)^{-b}},
\end{equation}
with $b\ge 1$. When written as a scalar-tensor theory, this is equivalent to a model with $n=-b/(1+b)$ \cite{Joyce:2014kja}. One can then see that 
Hu-Sawicki $f(R)$ models cover the narrow range of parameter space $-1<n<-1/2$. In particular, the most commonly studied models $b=1$ and $b=3$ 
correspond to $n=-1/2$ 
and $n=-3/4$ respectively. Other models, such as the designer model \cite{Song:2006ej,Nojiri:2006be,Nojiri:2006gh,Pogosian:2007sw,Nojiri:2010wj}, can 
span different ranges in $n$. This review focuses on general chameleon models and so we will not include any constraints that apply strictly to $f(R)$ 
models, although we will indicate interesting models on our final results. We refer the reader to \cite{Lombriser:2014dua} for a comprehensive review 
of $f(R)$ chameleons.

\section{Screening}

In this section we describe the behaviour of the chameleon in both astrophysical and laboratory settings.

\subsection{Astrophysical Screening}\label{sec:astroscreen}

Most astrophysical objects of interest are well approximated by spheres. When this is the case\footnote{The more general elliptical case has been 
studied by \cite{Burrage:2014daa}, who find that the screening is weakened by high ellipticity. }, the screening and force profile are captured by the 
fifth-force strength $\alpha$ and the self-screening parameter given by
\begin{align}
\alpha&=2\left(\frac{\mpl}{M}\right)^2\quad\textrm{and}\label{eq:astrodefs1}\\\chi&=\frac{\pmi(\rho_0)M}{2\mpl^2}=\frac{1}{2}\left(\frac{M}{\mpl^2} 
\right)^{\frac{n+2}{n+1}}\left(\frac{n\Lambda^{4+n}}{3\Omega_{{\rm m},0}H_0^2}\right)^{\frac{1}{n+1}},\label{eq:astrodefs2} 
\end{align}
where we have substituted $\rho_0=3\Omega_{{\rm m}\,0}\mpl^2H_0^2$ into equation \eqref{eq:pmi}. $\alpha$ parametrises the strength of the 
fifth-force and one has $F_5/F_{\rm N}=\alpha$ if the object is fully unscreened; $\alpha=1/3$ in $f(R)$ models. $\chi$ determines how 
efficient an object is at screening itself\footnote{the $f(R)$ literature often uses $f_{R0}=2/3\chi$ instead of $\chi$.}. In particular, $\chi$ determines the screening radius $\rs$ via the implicit relation
\begin{equation}\label{eq:screenrad}
\chi=4\pi G\int_{\rs}^Rr\rho(r)\dd r.
\end{equation}
Consider a theory with $\chi=0$ so that $\rs=R$. Increasing $\chi$ requires one to decrease $\rs$ (i.e. integrate further into the object) to satisfy this relation and therefore more of the object is unscreened. Theories with large $\chi$ therefore have a smaller screening radius and are hence more unscreened. If $\chi$ is so large that there is no solution then $\rs=0$ and the object is fully unscreened. One can show that this happens when $\chi>GM/Rc^2=\Phi_{\rm N}$, where $\pn$ is the surface Newtonian potential \cite{Davis:2011qf,Hui:2009kc,Sakstein:2013pda,Sakstein:2015oqa}. A good rule of thumb is then that objects are screened when $\chi<\pn$. 

One can see then that unscreened objects are those with low Newtonian potentials, which gives us a classification scheme for the level of self-screening. 
In particular, table \ref{tab:potentials} shows the Newtonian potential of commonly used astrophysical probes of chameleon models. Both the Sun and 
the Milky Way (being a spiral galaxy) have potentials of order $10^{-6}$ and so $\chi$ is constrained to lie below this value from the outset by the 
requirement that they are self-screening \footnote{In principle, one can relax this assumption by requiring the screening to be 
due to the local group instead. This is difficult to calculate and we will see shortly that astrophysical tests place constraints that are stronger 
than the requirement for self-screening and so we will assume self-screening from the outset.}. 

\begin{table}[h]\bgroup
\centering
\def\arraystretch{1.5}
 \begin{tabular}{c | c}
 Object & $\pn$ \\\hline
 Main-sequence star& $10^{-6}$\\
 Post-main-sequence star ($1$--$10M_\odot$) & $10^{-7}$--$10^{-8}$\\
 Spiral Galaxy & $10^{-6}$\\
 Dwarf Galaxy & $10^{-8}$
 \end{tabular}\caption{The Newtonian potential of useful astrophysical objects.}\label{tab:potentials}
  \egroup
\end{table}

Dwarf galaxies have low Newtonian potentials due to their slow rotation\footnote{Recall from the virial theorem that $GM/r\sim v^2$. Dwarf galaxies 
have rotational velocities of order $50$ km/s.}, which makes them perfect laboratories for testing chameleon theories. This has motivated 
\cite{Cabre:2012tq} to compile a screening map of nearby galaxies in the SDSS survey that gives the screening status of each galaxy as a function of 
$\chi$ accounting for environmental screening. Several tests using dwarf galaxies have been proposed \cite{Jain:2011ji} and we discuss these below. It is worth noting that there are no objects in the Universe with 
$\pn<10^{-8}$ that do not reside in screened galaxies. For this reason, there is a limit on the constraining power of astrophysical tests and models 
where $\chi<10^{-8}$ must be tested using other means.

\subsection{Laboratory Screening}

The precision laboratory measurements we consider here are performed in high vacuum. If the walls of the vacuum chamber are sufficiently thick, then the chameleon field can reach the minimum of the effective potential inside the wall, and the mass of the field increases. This means that the chameleon field in the interior of the vacuum chamber is effectively decoupled from its behaviour in the exterior and, in particular, is not affected by any environmental screening due to the Earth or Solar System. In this case, the equations governing the evolution of the field in the interior can be solved without any reference to the exterior solution. This condition is satisfied when the walls are thicker than $\sim 1/ m_{\rm eff}(\rho_{\rm wall})$. This also ensures that any chameleon signals propagating from the exterior towards the interior of the vacuum chamber are exponentially damped by the walls. 

If the vacuum chamber is large enough the chameleon field will take the value that minimises its effective potential in the centre. {In} smaller chambers the field does not have enough space to evolve from the field value that minimises the potential in the walls to that which minimises the potential in vacuum \cite{Khoury:2003rn}. In this case the background value of the field $\phi_0$ is set by the geometry of the vacuum chamber and the field takes a background value to ensure that its Compton wavelength is approximately the size of the chamber. For spherical vacuum chambers, this value can be calculated analytically \cite{Burrage:2014oza,Brax:2013cfa} but in more complicated scenarios it must be done numerically \cite{Elder:2016yxm,Schlogel:2015uea}. 

Once the background field value has been determined, the screening of sources can be computed.  Most sources used in laboratory experiments can be modeled to a good approximation as having constant density. In this case the expression for the screening radius simplifies and becomes
\begin{equation}
1-\frac{\rs^2}{R^2}= \left(\frac{M}{\mpl}\right)^2\frac{8 \pi \mpl^2 R}{M_{\rm obj}}\left(\frac{\phi_0-\phi_{\rm min}(\rho_{\rm obj})}{M}\right)
\label{eq:rsconst}
\end{equation}
The right hand side of Equation (\ref{eq:rsconst}) is the ratio of the chameleon to Newtonian potentials surrounding the source, multiplied by the square of the ratio of the chameleon and gravitational coupling strengths. 
 It has been demonstrated that, at least in parts of the chameleon parameter space, neutrons, atoms, and silica microspheres are unscreened, meaning that there is no real solution to Equation (\ref{eq:rsconst}) for these objects \cite{Burrage:2014oza,Rider:2016xaq}. 

There is an additional level of complexity when considering atomic and sub atomic particles, as measurements are performed with quantum states. However whilst the position of the particle may be uncertain, the particle or nucleus still has a well defined size as experiments are performed at low energies that do not disrupt the quantum chromodynamics binding the nucleons together. A trapped quantum particle will remain in a region of size $R_{\rm trap}$ for a time of order $R_{\rm trap}/v$, where $v$ is the deterministic velocity of the particle. The chameleon can respond to fluctuations in the position of the particle on a time scale set by $1/ m_{\rm eff}(\phi_0)$. When this is smaller than $R_{\rm trap}/v$, the chameleon can track the quantum fluctuations of the particle and whether the particle is screened should be determined from the size and mass of the nucleus or neutron. This is the case for all experiments that we consider here \cite{Burrage:2014oza}. 

\section{Experimental Tests of the Chameleon}\label{sec:experiments}

In this section we review various tests of chameleon dark energy models; we will combine them into one single set of constraints in section \ref{sec:constraints}.

\subsection{Astrophysical Tests}\label{sec:astro}

As remarked above, astrophysical tests typically focus on objects inside dwarf galaxies. Tests using other objects such as binary pulsars \cite{Brax:2013uh} and extra-solar planets \cite{Santos:2016rdg} are superseded. 

\subsubsection{Cepheid Tests}

Cepheid variable stars with masses between $4$ and $10M_\odot$ pulsate with a known period-luminosity relation and can hence be used as standard candles to measure the distance to other galaxies. The period scales as $\Pi\sim\sqrt{R^3/GM}$ and is hence sensitive to the theory of gravity. In particular, unscreened Cepheids pulsate with a shorter period leading one to underestimate the distance by a factor \cite{Jain:2012tn}
\begin{equation}
\frac{\Delta d}{d}\approx -0.3\frac{\Delta G}{G}.
\end{equation}
Another independent method to estimate the distance is to use the tip of the Red Giant branch (TRGB) \cite{Freedman:2010xv}. Post-main-sequence stars with masses $1M_\odot\lsim M\lsim2M_\odot$ do not exhibit instabilities but instead ascend the Red Giant branch burning hydrogen in a thin shell around their core. As this happens, the core temperature rises steadily until helium burning can proceed efficiently. This signals the onset of the so-called \emph{helium flash} where the
star rapidly moves onto the asymptotic giant branch (AGB) leaving a visible discontinuity in the I-band. This discontinuity occurs at fixed luminosity and is largely insensitive to the metallicity, making the TRGB a standard candle. Since the helium flash is set by nuclear and not gravitational physics this distance estimate is insensitive to chameleons.

\cite{Jain:2012tn} have used a sample of 25 galaxies from the screening map of \cite{Cabre:2012tq} to compare Cepheid and TRGB distance estimates to screened and unscreened samples of galaxies. Both samples are consistent with each other and the resulting $\chi^2$-fit to both GR and chameleon models places stringent bounds in the $\alpha$--$\chi$ plane, which we translate into our $n$--$\Lambda$--$M$ parameterisation using eqs. \eqref{eq:astrodefs1} and \eqref{eq:astrodefs2} in the next section.

\subsubsection{Rotation Curve Tests}

As mentioned above (and elaborated on in appendix \ref{sec:covtheory}), chameleon theories do not satisfy the equivalence principle; screened objects do not respond to chameleon fields whereas unscreened objects do. When $\chi<10^{-6}$ main-sequence stars in circular orbits around the galactic centre are screened and do not fell the fifth-force. In contrast, diffuse gas with a lower Newtonian potential is fully unscreened and feels the fifth-force in full. For completely unscreened galaxies where $F_5=\alpha F_{\rm N}$ the circular velocity law $v^2=F_{\rm N} + F_5$ implies that the circular velocity of the stars is larger than that of the gas by a factor
\begin{equation}
\frac{v_{\rm gas}}{v_\star}=\sqrt{1+\alpha}.
\end{equation}
An offset between the gaseous and stellar rotation curves then constitutes a novel test of chameleons.

Unfortunately, the galactic rotation curves are typically measured using H$\alpha$ or the 21 cm line, which both probe the gaseous component. The screening map of \cite{Cabre:2012tq} contained six unscreened low surface brightness galaxies for which information about the the Mgb triplet line was also available. This line is due to absorption in stellar atmospheres and hence allows the stellar rotation curve to be reconstructed. \cite{Vikram:2014uza} reconstructed the rotation curves for both components using this information and were able to place bounds in the $\alpha$--$\chi$ plane by looking at the confidence with which they could reject the predicted difference based on the observed difference on a galaxy-by-galaxy basis.

\subsubsection{Galaxy Cluster Tests}

The mass of galaxy clusters can be inferred using two independent methods. Hot, non-relativistic gas in the intra-cluster medium is in hydrostatic equilibrium and one can define a \emph{hydrostatic mass} via
\begin{equation}
\frac{\dd P}{\dd r} = -\frac{GM_{\rm hydro}(r)\rho}{r^2}.
\end{equation}
In GR $M_{\rm hydro}$ is the same as the cluster's mass $M$ (defined as the integral of the baryonic density) but in chameleon theories there is an additional contribution from the scalar force, which appears a correction to the hydrostatic mass. The pressure distribution is related to the X-ray surface brightness of the cluster, which allows the hydrostatic mass to be calculated from X-ray observations.

A second, independent mass measurement can be made using weak lensing. As explained in appendix \ref{sec:covtheory}, the lensing of light by massive objects is insensitive to chameleon gravity and therefore the mass inferred from weak lensing is identical to the true mass in both chameleon theories and GR \cite{Schmidt:2010jr}. Chameleon theories can therefore be probed by measuring and comparing the hydrostatic and lensing mass. \cite{Terukina:2013eqa} have performed such a test for the Coma cluster and, recently, \cite{Wilcox:2015kna} have applied the same technique to a sample of 58 clusters using X-ray data from the XMM Cluster Survey and lensing data from CFHTLenS to find new constraints.

\subsection{Fifth-force Searches}

Fifth-force searches look for new non-gravitational interactions by directly measuring the total force between different objects. Vacuum chambers are typically used to reduce noise and the geometry of the experiments is chosen to minimise the Newtonian force. In many cases the experiments probe length-scales below the dark energy scale of $\sim 90\mu\textrm{m}$.

\subsubsection{Torsion Balance Experiments}\label{sec:eotwash}

Torsion balance experiments probe the inverse-square law and are therefore excellent probes of massive scalars and chameleons due to their 
Yukawa-like force laws (see equation \eqref{eq:f5}). In particular, the E\"{o}t-Wash experiment 
\cite{Adelberger:2003zx,Kapner:2006si,Adelberger:2006dh} has been used to probe fifth-forces on sub-mm scales. The experiment consists of a torsion 
pendulum in the form of a freely rotating circular disk suspended above a second disk that acts as an attractor. Both disks have holes drilled into 
them that act as missing masses producing dipole and higher-order multipole moments. These cause a torque on the pendulum when the holes are 
misaligned. The attractor is rotated at a uniform angular velocity chosen such that there is no torque on the pendulum if the force is exactly 
inverse-square. Conversely, deviations from this form predict large torques, the absence of which can be used to constrain the model parameters. 

Both the pendulum and the attractor are coated in gold to reduce electrostatic effects and a beryllium-copper membrane is placed between the pendulum 
and the attractor for the same purpose. These have little effects for linear modifications of gravity such as Yukawa interactions but cause 
considerable complications when one wants to make theoretical predictions for non-linear chameleon models. In particular, the gold coatings and BeCu 
membrane may either have a thin shell or not depending on the model parameters. Furthermore, the non-linear nature of the field equations makes 
the computation of the field profile difficult for non-symmetric systems such as the E\"{o}t-Wash configuration. This has led to an intense 
effort towards making ever-increasingly precise approximations for the chameleon torque 
\cite{Mota:2006ed,Mota:2006fz,Adelberger:2006dh,Brax:2008hh,Upadhye:2012fz,Upadhye:2012qu}. The most stringent constraints have been found using the 
so-called ``one-dimensional plane-parallel approximation \cite{Upadhye:2012qu}, which attempts to include the effects of the missing masses on the 
field profile. 

\subsubsection{Casimir Force Tests}\label{sec:casimir}

Quantum electrodynamics predicts a force between two conducting plates due to the exchange of virtual photons that has been measured to 
incredibly high precision \cite{Lamoreaux:1996wh,Lambrecht:2011qm}. The Casimir force per unit area between two parallel plates is proportional to 
$d^{-4}$ where $d$ is the plate separation. When the plates have thin shells, the chameleon force per unit area scales as 
\cite{Mota:2006fz,Brax:2007vm,Brax:2014zta}
\begin{equation}
 \frac{F_\phi}{A}\sim d^{-\frac{2n}{n+2}},
\end{equation}
which always scales with a power $\ge-4$ with equality when $n=-4$. Thus, the chameleon force dominates at large separations and can be probed by the 
agreement of the measured force with the predicted one. In practice, parallel plates are inconvenient for measurements since they are difficult to 
keep parallel to the required precision and very smooth plates are required. Instead, the most precise experiments measure the force between a plate 
and a sphere whose radius of curvature is large compared with the minimum separation. In this case, the total Casimir force is proportional to 
$d^{-3}$ while the chameleon force scales as
\begin{equation}
 F_\phi\sim d^{\frac{2-n}{n+2}},
\end{equation}
which again always scales with a power $\ge -3$ and hence the same principles apply. 

Current Casimir force searches \cite{Decca:2007yb} place strong bounds on $n=-4$ and $n=-6$ models when $\Lambda$ is dark energy-scale 
\cite{Brax:2007vm} but are not competitive with other probes when $n>0$. Planned future experiments \cite{Lambrecht:2005km,Lamoreaux:2005zza} 
that propose to use larger separations will place more stringent constraints. Finally, the non-linear nature of the chameleon allows for more 
targeted experiments. For example, by holding the plates still and varying the pressure of the ambient gas the chameleon could be probed by looking 
for the characteristic change of force with gas density \cite{Brax:2010xx,Almasi:2015zpa}. 

\subsubsection{Levitated Microspheres}

Optically levitated dielectric spheres with radii $\sim\mathcal{O}(\mu\textrm{m})$ can be used to probe forces $\lsim\mathcal{O}(10^{-8}\textrm{N})$ \cite{Geraci:2010ft}. Microspheres can be unscreened when $\Lambda\gsim4.6$ meV and recently \cite{Rider:2016xaq} have used this to constrain chameleon forces. The radiation pressure of a single upward-pointing laser beam trap acts to counter the Earth's gravity so that any additional forces dominate. These are measured using a cantilever and in the case of chameleons are given by 
\begin{equation}
F=\lambda\left(\frac{\rho}{M}\right)\int\dd^3\vec{x}\frac{\partial\phi}{\partial z},
\end{equation}
 where the integral is performed over the volume of the sphere (the density $\rho$ is constant) and $z$ is the vertical direction. The scalar charge $\lambda$ is unity when the sphere is unscreened, which is the case for $M\lsim 10^{10}$ TeV, but at larger $M$ screening reduces this so that $\lambda<1$, limiting the sensitivity. Currently, only constraints on $n=1$ theories have been reported \cite{Rider:2016xaq}.

\subsection{Atom Interferometry}\label{sec:atomint}

Atom interferometry is a technique used to measure forces on individual atoms. This is beneficial for the chameleon because over a wide range of 
the chameleon parameter space atoms are unscreened in a laboratory vacuum. The experiment can be thought of as a cross between a classical 
Michaelson-Morley interferometer and the double slit experiment of quantum mechanics.  An atom is put into a superposition of states which travel 
along different paths, these are the two arms of the interferometer. The two paths are then recombined and a measurement is made. The atoms are moved 
around in the interferometer by shining laser beams at them; if the atom absorbs a photon, exciting an electron to a higher energy state, then 
the atom also absorbs the photon's momentum and thus acquires some linear motion. If no observation is made at this point the atom is in a 
superposition of the ground state where the atom is stationary and an excited state where the atom is moving. By repeating this process it is 
possible to put the atom into a superposition of states that travel along the two arms of the interferometer. When the two paths are recombined the 
probability that the atom is observed to be in its excited state is 
\begin{equation}
P \propto 2 \cos \left[\frac{a T^2 k}{\hbar}\right]
\end{equation}
where $a$ is the acceleration (assumed constant) experienced by the atoms, $k$ is the momentum of the photons and $2T$ is the duration of the experiment.

If a massive source is placed inside the vacuum chamber then the atoms experience an acceleration towards due to the chameleon force 
\cite{Burrage:2014oza,Burrage:2015lya}. It has recently been possible to constrain this acceleration using atom interferometry down to a precision of 
$10^{-6} g$, where $g$ is the acceleration due to free fall at the surface of the Earth \cite{Hamilton:2015zga,Elder:2016yxm}. It is expected that sensitivities of 
$10^{-9}g$ could be reached in the near future. 

\subsection{Precision Atomic tests}\label{sec:atomic}
The chameleon mediates an attractive fifth-force between the nuclei of atoms and their electrons. This will perturb the motion of the electrons from 
the motion that would be calculated using only the Coulomb potential. As described above, the walls of a vacuum chamber screen the interior of the 
chamber from any chameleon effects in the exterior, and therefore strongly coupled chameleons could produce observable shifts in the energy levels of 
atomic nuclei \cite{Brax:2010gp}. 

The Hamiltonian describing the electrons orbiting an hydrogenic atom is perturbed by a term
\begin{equation}
\delta H = \frac{m_e}{M}\phi_{\rm nuc}
\end{equation}
where $m_e$ is the mass of the electron, and $\phi_{\rm nuc}$ is the chameleon field profile sourced by the atomic nucleus. Assuming that the nucleus 
is sufficiently light to be unscreened inside the vacuum chamber, the 1s, 2s, and 2p energy levels are perturbed by
\begin{align}
\delta E_{1s} =& -\frac{Zm_Nm_e}{4 \pi M^2 a_0}\\
\delta E_{2s}= \delta E_{2p} =& -\frac{Zm_Nm_e}{16 \pi M^2 a_0}
\end{align}
where $Ze$ is the nuclear charge, $m_N$ the mass of a nucleon and $a_0$ the Bohr radius. If a coupling of the chameleon to photons is introduced the 
degeneracy between the 2s and 2p energy levels will be broken.

The best current measurement for constraining the chameleon is that of the 1s-2s transition for a hydrogen atom. This measurement has a total 
uncertainty of $10^{-9} \mbox{ eV}$ at $1\sigma$ \cite{Jaeckel:2010xx,Schwob:1999zz,Simon:1980hu}, and agrees with the prediction of the standard 
Coulombic calculation. This therefore constrains the perturbation due to the chameleon to be smaller than $10^{-9} \mbox {eV}$ which implies
\begin{equation}
M\gtrsim 10 \mbox{ TeV}.
\end{equation}

\subsection{Precision Neutron Constraints}

Being electrically neutral, neutrons provide a new window into low-energy particle physics and short-range gravitational physics precisely because 
they are free from the electromagnetic backgrounds, van der Waals interactions, and Casimir-Polder forces that act as a source of uncertainty for charged 
particle experiments. Several different experiments using slow neutrons have placed competitive constraints on the coupling $M$ \cite{Brax:2013cfa}, 
which we summarise below. The constraints were found assuming that $\Lambda=2.4\times10^{-3}$ eV (i.e. the dark energy scale) in all cases.

\subsubsection{Ultra Cold Neutrons}\label{sec:UCN}

Ultra cold neutrons bouncing above a mirror interact with the gravitational potential of the Earth leading to a quantised energy spectrum. The 
presence of a chameleon field sourced by the mirror $\phi(z)$, where $z$ is the distance above the mirror, introduces a perturbation to the Hamiltonian 
\cite{Brax:2011hb,Ivanov:2012cb}
\begin{equation}
 \delta {H} = \frac{ m_{\rm N}}{M}\phi(z)= \frac{2.2 \textrm{keV}^2}{M}\left(\frac{z}{\textrm{82}\,\mu\textrm{m}}\right)^{\frac{2}{2+n}},
\end{equation}
where $m_{\rm N}$ is the neutron mass. If this were too large, new bound states not seen by the Grenoble experiments \cite{Nesvizhevsky:2002ef} would 
exist, which places the new bound \cite{Brax:2011hb}
\begin{equation}
 M > 10^4 \textrm{ TeV}. 
\end{equation}
Furthermore, the Hamiltonian perturbs the energy levels, which can be measured using resonance spectroscopy of the $|3\rangle\rightarrow |1\rangle$ 
transition \cite{Jenke:2014yel} to find the stronger bound
\begin{equation}
 M > 1.7\times10^6 \textrm{ TeV}. 
\end{equation}

\subsubsection{Neutron Interferometry}\label{sec:neutronint}

Neutron interferometry measures the quantum properties of neutrons by splitting a monochromatic beam into two coherent beams using a mono-silicone 
crystal plate. The beam is later recombined using similar plates to produce an interference pattern whose phase shift can be extracted 
\cite{Brax:2013cfa}. If one of the beams traverses a vacuum chamber with walls located at $x=\pm R$ an additional phase shift 
\begin{equation}
 \delta\varphi=\frac{\sqrt{2} m_{\rm N}^2 K_{n}(0)}{\hbar^2k M}\left(\frac{\sqrt{2}R\Lambda}{J_n(0)}\right)^{\frac{n+4}{n+2}},
\end{equation}
where $J_n$ and $K_n$ are Bessel functions of the first and second kind, arises due to the chameleon field between the plates 
\cite{Brax:2013cfa,Brax:2014gja}. This experiment has been performed by \cite{Lemmel:2015kwa,Li:2016tux} who report the bounds listed in table \ref{tab:neutint}.
\begin{table}[h]\bgroup
\centering
\def\arraystretch{1.5}
 \begin{tabular}{c| c | c}
 Model & Reference \cite{Lemmel:2015kwa} & Reference \cite{Li:2016tux} \\\hline
 $n=1$ & $M > 5.3\times10^7$ TeV & $M>2.1\times10^8$ TeV \\
 $n=2$ & $M > 1.7\times10^7$ TeV &$M>1.2\times10^8$ TeV\\
 $n=3$ & $M > 5.0\times10^6$ TeV &$M>7.9\times10^7$ TeV \\
 $n=4$& $M > 2.1\times10^6$ TeV &$M>5.6\times10^7$ TeV\\
 $n=5$ & - &$M>4.9\times10^7$ TeV\\
 $n = 6$ & - &$M>4.2\times10^7$ TeV
 \end{tabular}\caption{The bounds on $M$ coming from neutron interferometry experiments \cite{Lemmel:2015kwa}.}\label{tab:neutint}
  \egroup
\end{table}

\subsection{Coupling to Photons}
We briefly mention a class of searches for chameleon fields that are orthogonal to those we consider here. These rely on introducing a coupling between the chameleon and photons, making the chameleon an axion-like particle. Such a coupling is not present in the simplest chameleon models because the photon has a traceless energy-momentum tensor but it is not forbidden and can be introduced into the model by hand or by integrating out charged, massive fields \cite{Brax:2009ey,Brax:2010uq}. Searches for such fields are closely related to searches for axion-like particles, relying on the Primakov 
effect to convert photons into chameleons (and vice versa) in the presence of a magnetic field. 

Terrestrial searches have been performed by Gammev-ChASe \cite{Steffen:2010ze}, ADMX \cite{Rybka:2010ah}, and CAST \cite{Anastassopoulos:2015yda}, and astrophysical observations of the polarization and luminosity of light from stars \cite{Burrage:2008ii} and active galactic nuclei \cite{Burrage:2009mj,Pettinari:2010ay} have all been used to constrain the strength of the chameleon--photon coupling. The chameleon mechanism means that such fields avoid many constraints on axion-like particles that are derived from dense environments, however the coupling to photons is not an intrinsic part of the chameleon mechanism, and constraining it does not tell us how the field changes its mass with the environment. Therefore we restrict our attention to the necessary couplings to massive fields in this work. 

\clearpage
\section{Combined Constraints}\label{sec:constraints}

\begin{figure*}[ht]
\begin{center}
\includegraphics[width=0.49\textwidth]{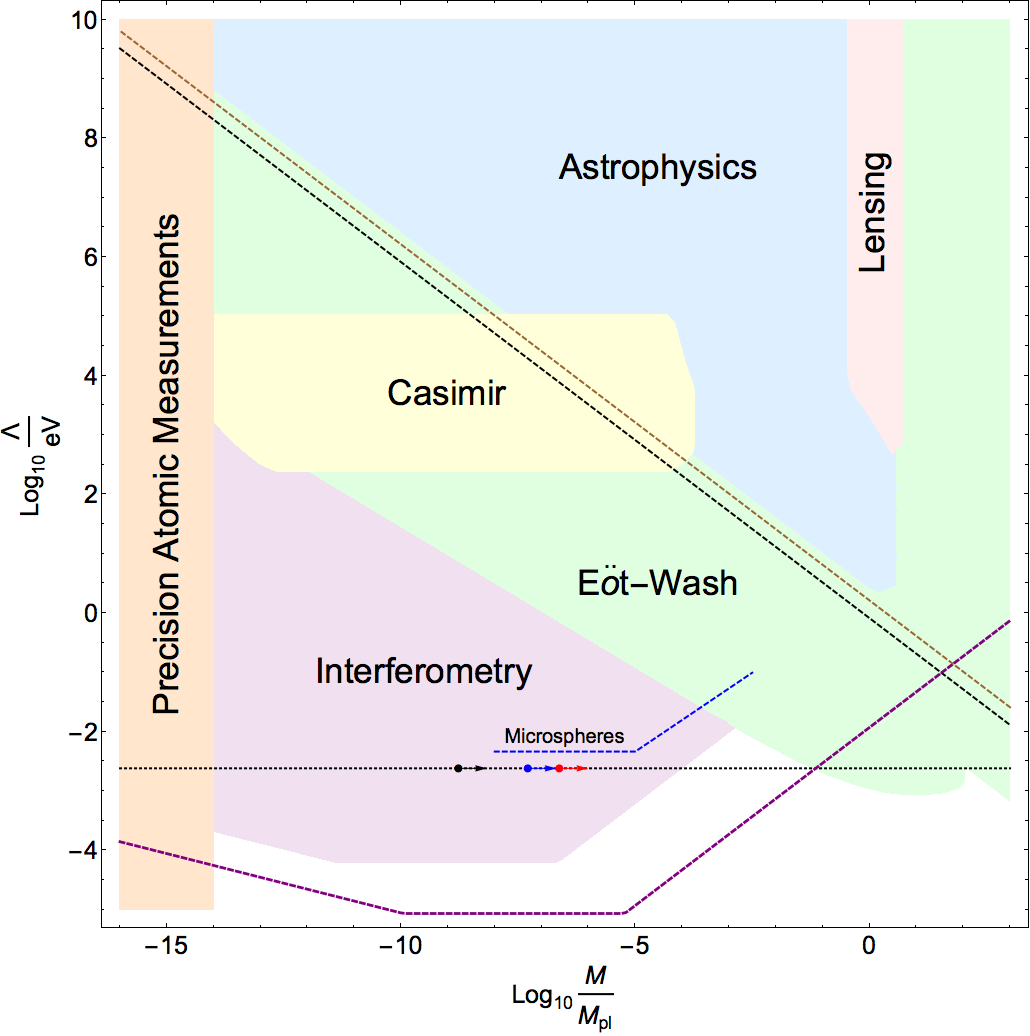}
\includegraphics[width=0.49\textwidth]{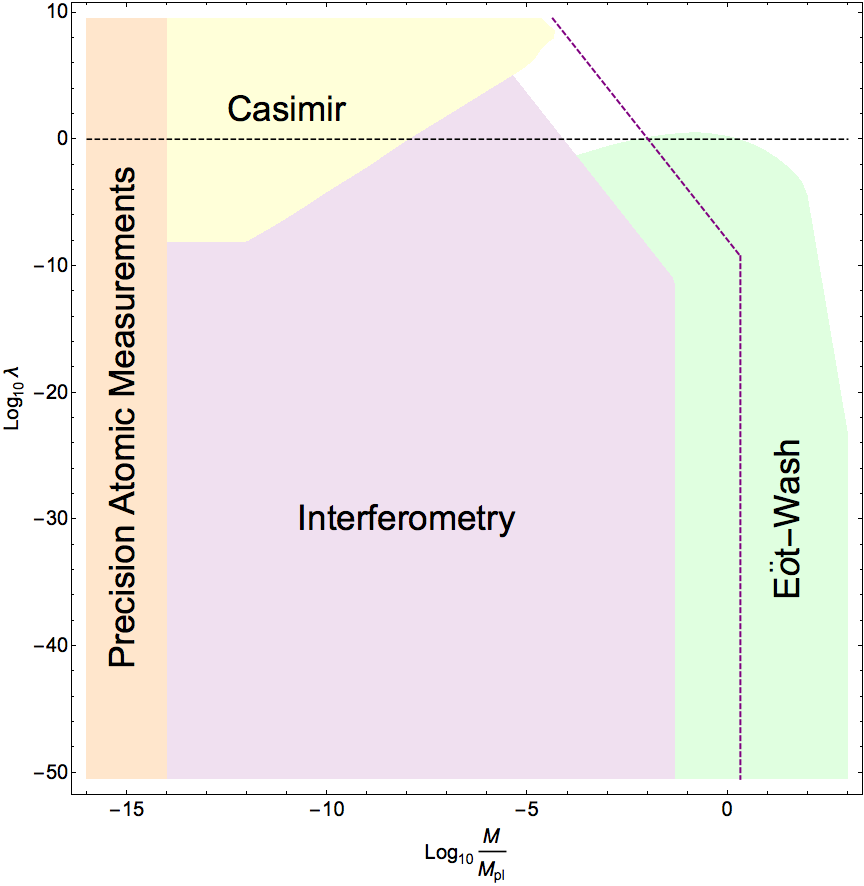}
\vspace{5mm}
\includegraphics[width=0.49\textwidth]{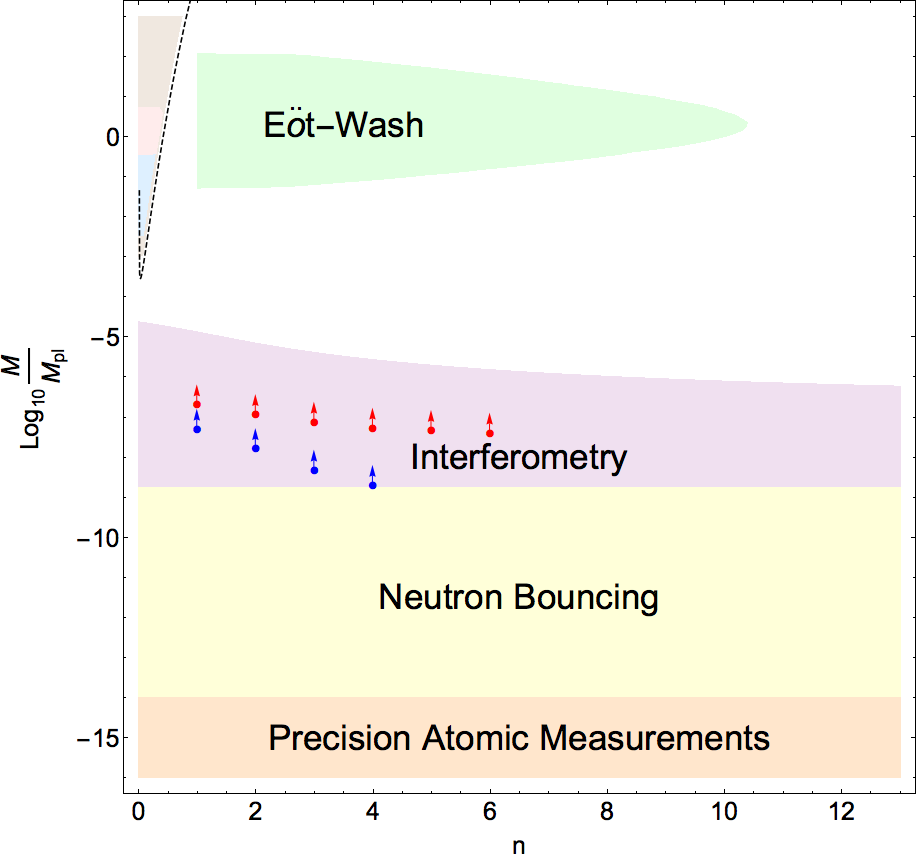}
\includegraphics[width=0.49\textwidth]{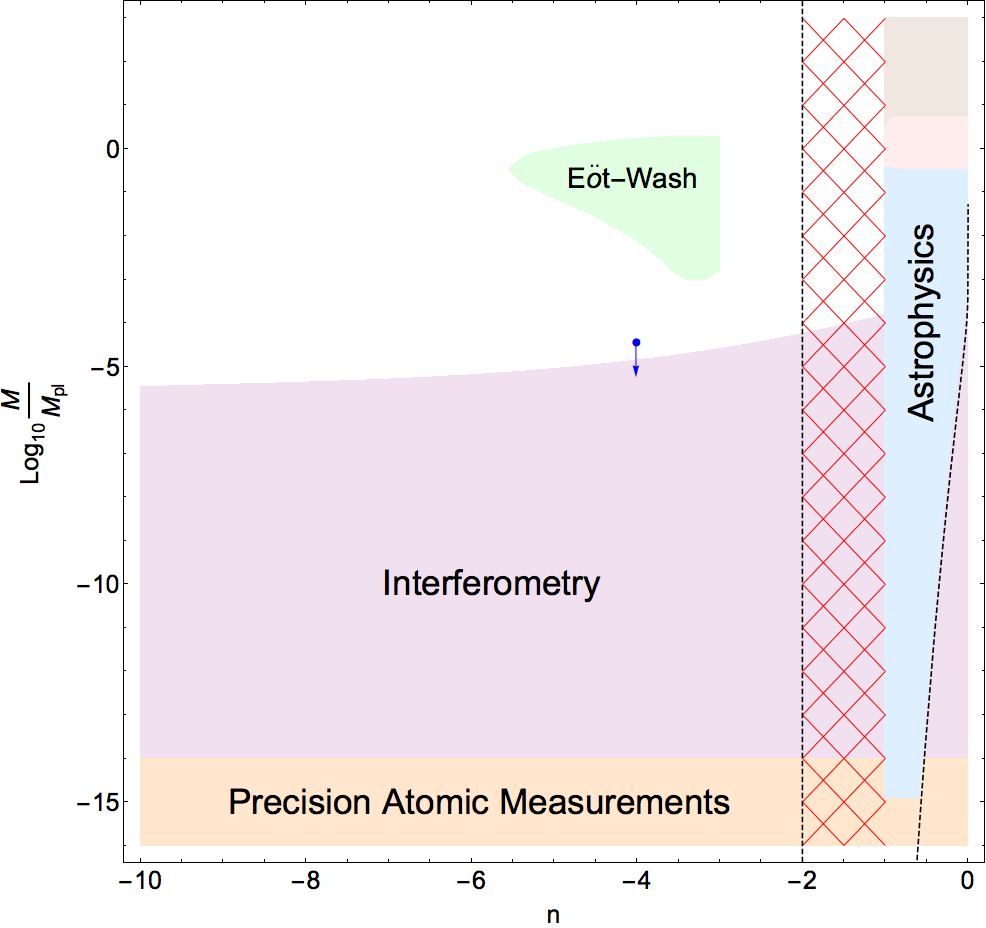}
\caption{Combined constraints on chameleon models for $n=1$ (top left panel), $n=-4$ (top right panel), and general $n$ with $\Lambda$ fixed to the dark energy scale, $2.4$ meV (positive $n$ in bottom left panel and negative $n$ in bottom right panel). The excluded regions from different experiments are labelled accordingly; rotation curve and Cepheid tests are combined in the blue astrophysics region and the pink region corresponds to cluster lensing constraints. The black, blue and red dots show the lower bounds (indicated by the arrow) on $M$ at the dark energy scale coming from neutron bouncing and interferometry experiments respectively, with blue corresponding to the bounds of \cite{Lemmel:2015kwa} and red to those of \cite{Li:2016tux}. The dark energy scale $\Lambda=2.4\times10^{-3}$ eV (upper plots only) and the astrophysical limit are indicated by the dotted and dashed black lines respectively. The dashed purple black line indicates the forecasted constraints from the next generation of atom interferometry experiments taken from \cite{Hamilton:2015zga} ($n=1$ and $n=-4$). The red hashed area (bottom right only) corresponds to regions where the model is not a chameleon. The brown region is accessible to cosmological observations. Note that in the case $n=1$ this region is already excluded by other probes and so we indicate it using the brown dashed line; the region above and to the right of this predicts observable deviations on cosmological scales. Whilst constraints for negative $n$ are plotted with continuous lines we remind the reader that only negative even integers give rise to a chameleon mechanism.} \label{fig:cons}
\end{center}
\end{figure*}

We have collated the most stringent results from the various probes described in section \ref{sec:experiments} and have translated them into the common paramterisation given by $\{n,M,\Lambda\}$ or, in the case $n=-4$, $\{\lambda,M\}$. We list the constraints we use and their associated references in table \ref{tab:constraints}. In many cases, the constraints apply only for specific values of $n$, $\Lambda$, or $\lambda$. In figure \ref{fig:cons} we show constraints for the $n=1$ and $n=-4$ models, which are the most commonly studied, and constraints in the $n$--$M$ plane for positive and negative $n$ separately with $\Lambda$ fixed to the dark energy scale. If $\Lambda_0=2.4\times10^{-3}$ eV is the dark energy scale then comparing with the case of arbitrary $n\ne-4$ one has $\lambda=(\Lambda/\Lambda_0)^4$ and so the range of $\lambda$ covers the same range of $\Lambda$ plotted above when $n=1$. Furthermore, the reader should recall that $n=-1$ and $n=-2$ models are not chameleons. Similarly, only negative even integers and those where $-1<n<0$ exhibit the mechanism and so the constraints in these plots should be interpreted as being valid for these specific values of $n$ only.

\begin{table}[h]\bgroup
\centering
\def\arraystretch{1.5}
 \begin{tabular}{c | c  }
 Experiment & Reference \\\hline
 Cepheids & \cite{Jain:2012tn} \\
 Rotation Curves & \cite{Vikram:2014uza} \\
 Cluster Lensing & \cite{Wilcox:2015kna} \\
 E\"{o}t-Wash & \cite{Upadhye:2012qu} \\
  Casimir Force Tests & \cite{Brax:2007vm} \\
  Microspheres & \cite{Rider:2016xaq}\\
 Atom Interferometry & \cite{Hamilton:2015zga} \\
 Precision Atomic Tests & \cite{Brax:2010gp} \\
 Neutron Bouncing & \cite{Jenke:2014yel} \\
  Neutron Interferometry & \cite{Lemmel:2015kwa,Li:2016tux}
 \end{tabular}\caption{The constraints used in this section.}\label{tab:constraints}
  \egroup
\end{table}

\clearpage
\section{Discussion}

We have shown that the majority of the chameleon parameter space is already excluded for models with $n=1$ and $n=-4$, and that if we fix $\Lambda$ to the cosmological constant scale then only the chameleon theories most weakly coupled to matter are allowed. Astrophysical constraints are complementary to laboratory searches but currently their relevance is limited to relatively small regions of the chameleon parameter space, {although} we note that these are {precisely} the regions where the chameleon {has a cosmological scale Compton wavelength.}

To date, there have been no constraints coming from cosmological probes but this will change with upcoming cosmological surveys such as EUCLID \cite{Amendola:2012ys}. One important question is then whether cosmological constraints will probe unexplored areas of the parameter space? And if so, which models are best constrained? To address these, we consider the equation for the growth of density perturbations in the Newtonian gauge \cite{Brax:2012gr,Brax:2013yja}
\begin{align}
0&=\ddot{\delta}+2H\dot{\delta}-\frac{3}{2}\Omega\mmm(a)H^2\frac{G\eff}{G}\delta=0\quad\textrm{with}\\\quad\frac{G\eff}{G}&={1+2\left(\frac{\mpl}{M}\right)^2}\left({1+\frac{a^2}{\lambda_{\rm C}^2k^2}}\right)^{-1},
\end{align}
where $\lambda_{\rm C}=m\eff^{-1}$ is the Compton wavelength (at cosmological densities). One can see that on large scales where $k\lambda_{\rm C}\ll1$ one has $G\eff\approx G$ and the force is screened, whereas on small scales $k\lambda_{\rm C}\gg1$ one has $G\eff\approx G(1+2\mpl^2/M^2)$ and the force is fully unscreened. Said another way, deviations from GR are only present on scales $k^{-1}<\lambda_{\rm C}$ i.e. scales inside the Compton wavelength. Linear cosmological probes will therefore see deviations from GR if $\lambda_{\rm C}\gsim 10 \textrm{Mpc}$. We have included the cosmologically viable regions in the constraint plots (figure \ref{fig:cons}). The $n=1$ models that are accessible cosmologically are already excluded by other methods and so we indicate the region with a brown dashed line. The same is true of $n>1$ when $\Lambda$ is fixed to the dark energy scale but models with $n<1$ may be probed. Similarly, there is a small region of parameter space where models with $-1<n<0$ can be probed with cosmology. In the case $n=-4$ the Compton wavelength is $\lsim\oo(10^{-7}\textrm{ Mpc})$ and so the region of parameter space we are interested in cannot be probed using cosmology. It is interesting to note that in all cases the edge of the cosmologically accessible region is close to the astrophysical limit. This is because one has $\chi\approx H^2/m\eff^2$ \cite{Upadhye:2012qu,Brax:2013yja,Brax:2012mq} so that the astrophysical limit $\chi=10^{-8}$ corresponds to $\lambda_{\rm C}\sim 10$ Mpc. One could therefore exclude the entire region where cosmology can probe these models by improving the astrophysical constraints, in particular extending the constraints to cover larger values of $M$ (weaker matter couplings).

Indeed, the astrophysical probes presented here are far from exhausted and are mainly limited by the small number of dwarfs in the screening map. Future SDSS data releases, in particular MaNGA, could drastically increase the sample size and with it, the constraints. On a similar note, other proposed tests using the morphology and kinematics of dwarf galaxies \cite{Jain:2011ji} have yet to be performed but may yield new constraints. \cite{Vikram:2013uba} considered several tests along these lines using SDSS optical and ALFALFA radio data but were unable to distinguish between modified gravity and GR due to the large scatter. Future radio observations such as VLA may improve the constraining power of such tests.

{From Figure \ref{fig:cons} it is clear that large regions of the chameleon parameter space are likely to remain inaccessible to cosmological and astrophysical probes for the foreseeable future. This, therefore, provides strong motivation for further development of laboratory searches for chameleon fields. We are aware of ongoing developments of Casimir, atom and neutron interferometry, and microsphere searches which should yield new, improved constraints in the near future, further restricting the allowed behaviour of the chameleon field.}

{The chameleon model of dark energy is now severely constrained by a wide variety of observations.  It seems very likely that we currently have the technological capability to perform measurements to detect or completely exclude the chameleon field wherever it lies in the parameter space.  All that remains is to perform the experiments.
}

\appendix

\section{Covariant Theory}\label{sec:covtheory}

Chameleons are a class of scalar-tensor theories that correspond to the action 
\begin{align}
S&\nonumber=\int\dd^4 x\sqrt{-g}\left[\frac{\mpl^2}{2}R(g)-\frac{(\nabla_\mu\phi)^2}{2}-V(\phi)\right]\\&+S_{\rm matter}[A^2(\phi)g_\nm];\quad A(\phi)=e^{\frac{\phi}{M}}
\end{align}
defined in the so-called Einstein frame where the tensor-sector looks like general relativity but the scalar is non-minimally coupled to matter through the coupling function $A(\phi)$. In particular, matter moves on geodesics of the \emph{Jordan frame} metric $\tg_\nm=A^2(\phi)g_\nm$. The equation of motion for the scalar is
\begin{equation}
\Box\phi=\frac{\dd V(\phi)}{\dd\phi}-T\frac{\dd\ln A(\phi)}{\dd\phi},
\end{equation}
where $T=g_\nm T^\nm$ is the trace of the energy-momentum tensor for matter $T^\nm=2/\sqrt{-g}\delta S\mmm/\delta g_\nm$. For non-relativistic systems ($P\ll\rho$ where $P$ is the pressure) one has $T=-\rho$ where $\rho$ is the matter density, in which case one recovers equation \eqref{eq:chameomrel}. 

The fifth-force can be found by noting that test-bodies move on geodesics of $\tg_\nm$. Defining the tensor $\mathcal{K}^\alpha_\nm=\tilde{\Gamma}^\alpha_\nm-\Gamma^\mu_\nm$ where $\Gamma$ are the Christoffel symbols and tildes refer to Jordan frame quantities, the non-relativistic limit of the geodesic equation in the Jordan frame becomes
\begin{equation}
\ddot{x}^i+\Gamma_{00}^i=-\mathcal{K}^i_{00},
\end{equation}
where a dot denotes a derivative with respect to $t$, the proper-time for an observer. One has the well-known result that $\Gamma_{00}^i=\partial^i\pn$, which is the Newtonian force in the Einstein frame and one can calculate $\mathcal{K}^i_{00}=\partial^i\phi$ in a straight-forward manner (see \cite{wald2010general,Brax:2012gr,Sakstein:2014isa,Sakstein:2015oqa,Sakstein:2015jca} for example). This then represents a fifth-force given by equation \eqref{eq:force}. In the case where the object is not a test mass but is instead an extended body one must self-consistently calculate the force using the method of \cite{Einstein:1938yz,Damour:1986ny}. This has been done by \cite{Hui:2009kc} who find that the equation of motion for a body of mass $M_{\rm obj}$ is
\begin{equation}\label{eq:EP}
M_{\rm obj}\vec{\ddot{x}}=-M_{\rm obj}\nabla\pn-Q\nabla\phi,
\end{equation}
where the scalar charge
\begin{equation}
Q=M_{\rm obj}-M_{\rm obj}(\rs)
\end{equation}
and we remind the reader that $M_{\rm obj}(\rs)$ is the mass enclosed within the screening radius. Equation \eqref{eq:EP} shows how the equivalence principle is violated in chameleon theories: two objects with identical masses but different internal compositions will have different screening radii by virtue of \eqref{eq:screenrad} and therefore have different scalar charges. Said another way, an object's response to an applied external field is not determined solely by its mass.

We end by noting that null geodesics of $\tg$ are also null geodesics of $g$ with a different affine parameter (see \cite{Padmanabhan:2010zzb} for example) and therefore the lensing of light is unaltered in chameleon theories. 

\bibliographystyle{jhep}
\bibliography{ref}

\providecommand{\href}[2]{#2}\begingroup\raggedright\begin{thebibliography}{10}

\bibitem{Khoury:2003aq}
J.~Khoury and A.~Weltman, {\it {Chameleon fields: Awaiting surprises for tests
  of gravity in space}},  {\em Phys. Rev. Lett.} {\bf 93} (2004) 171104,
  [\href{http://arxiv.org/abs/astro-ph/0309300}{{\tt astro-ph/0309300}}].

\bibitem{Khoury:2003rn}
J.~Khoury and A.~Weltman, {\it {Chameleon cosmology}},  {\em Phys. Rev.} {\bf
  D69} (2004) 044026, [\href{http://arxiv.org/abs/astro-ph/0309411}{{\tt
  astro-ph/0309411}}].

\bibitem{Adelberger:2003zx}
E.~G. Adelberger, B.~R. Heckel, and A.~E. Nelson, {\it {Tests of the
  gravitational inverse square law}},  {\em Ann. Rev. Nucl. Part. Sci.} {\bf
  53} (2003) 77--121, [\href{http://arxiv.org/abs/hep-ph/0307284}{{\tt
  hep-ph/0307284}}].

\bibitem{Adelberger:2005vu}
E.~Adelberger, B.~R. Heckel, and C.~D. Hoyle, {\it {Testing the gravitational
  inverse-square law}},  {\em Phys. World} {\bf 18N4} (2005) 41--45.

\bibitem{Kapner:2006si}
D.~J. Kapner, T.~S. Cook, E.~G. Adelberger, J.~H. Gundlach, B.~R. Heckel, C.~D.
  Hoyle, and H.~E. Swanson, {\it {Tests of the gravitational inverse-square law
  below the dark-energy length scale}},  {\em Phys. Rev. Lett.} {\bf 98} (2007)
  021101, [\href{http://arxiv.org/abs/hep-ph/0611184}{{\tt hep-ph/0611184}}].

\bibitem{Murphy:2012rea}
T.~W. Murphy, Jr., E.~G. Adelberger, J.~B.~R. Battat, C.~D. Hoyle, N.~H.
  Johnson, R.~J. McMillan, C.~W. Stubbs, and H.~E. Swanson, {\it {APOLLO:
  millimeter lunar laser ranging}},  {\em Class. Quant. Grav.} {\bf 29} (2012)
  184005.

\bibitem{Hui:2009kc}
L.~Hui, A.~Nicolis, and C.~Stubbs, {\it {Equivalence Principle Implications of
  Modified Gravity Models}},  {\em Phys. Rev.} {\bf D80} (2009) 104002,
  [\href{http://arxiv.org/abs/0905.2966}{{\tt arXiv:0905.2966}}].

\bibitem{Chang:2010xh}
P.~Chang and L.~Hui, {\it {Stellar Structure and Tests of Modified Gravity}},
  {\em Astrophys. J.} {\bf 732} (2011) 25,
  [\href{http://arxiv.org/abs/1011.4107}{{\tt arXiv:1011.4107}}].

\bibitem{Davis:2011qf}
A.-C. Davis, E.~A. Lim, J.~Sakstein, and D.~Shaw, {\it {Modified Gravity Makes
  Galaxies Brighter}},  {\em Phys. Rev.} {\bf D85} (2012) 123006,
  [\href{http://arxiv.org/abs/1102.5278}{{\tt arXiv:1102.5278}}].

\bibitem{Sakstein:2013pda}
J.~Sakstein, {\it {Stellar Oscillations in Modified Gravity}},  {\em Phys.
  Rev.} {\bf D88} (2013), no.~12 124013,
  [\href{http://arxiv.org/abs/1309.0495}{{\tt arXiv:1309.0495}}].

\bibitem{Sakstein:2015oqa}
J.~Sakstein, {\em {Astrophysical Tests of Modified Gravity}}.
\newblock PhD thesis, Cambridge U., DAMTP, 2014.
\newblock \href{http://arxiv.org/abs/1502.0450}{{\tt arXiv:1502.0450}}.

\bibitem{Wang:2012kj}
J.~Wang, L.~Hui, and J.~Khoury, {\it {No-Go Theorems for Generalized Chameleon
  Field Theories}},  {\em Phys. Rev. Lett.} {\bf 109} (2012) 241301,
  [\href{http://arxiv.org/abs/1208.4612}{{\tt arXiv:1208.4612}}].

\bibitem{Brax:2011aw}
P.~Brax, A.-C. Davis, and B.~Li, {\it {Modified Gravity Tomography}},  {\em
  Phys. Lett.} {\bf B715} (2012) 38--43,
  [\href{http://arxiv.org/abs/1111.6613}{{\tt arXiv:1111.6613}}].

\bibitem{Brax:2004qh}
P.~Brax, C.~van~de Bruck, A.-C. Davis, J.~Khoury, and A.~Weltman, {\it
  {Detecting dark energy in orbit - The Cosmological chameleon}},  {\em Phys.
  Rev.} {\bf D70} (2004) 123518,
  [\href{http://arxiv.org/abs/astro-ph/0408415}{{\tt astro-ph/0408415}}].

\bibitem{Upadhye:2012vh}
A.~Upadhye, W.~Hu, and J.~Khoury, {\it {Quantum Stability of Chameleon Field
  Theories}},  {\em Phys. Rev. Lett.} {\bf 109} (2012) 041301,
  [\href{http://arxiv.org/abs/1204.3906}{{\tt arXiv:1204.3906}}].

\bibitem{Erickcek:2013oma}
A.~L. Erickcek, N.~Barnaby, C.~Burrage, and Z.~Huang, {\it {Catastrophic
  Consequences of Kicking the Chameleon}},  {\em Phys. Rev. Lett.} {\bf 110}
  (2013) 171101, [\href{http://arxiv.org/abs/1304.0009}{{\tt
  arXiv:1304.0009}}].

\bibitem{Erickcek:2013dea}
A.~L. Erickcek, N.~Barnaby, C.~Burrage, and Z.~Huang, {\it {Chameleons in the
  Early Universe: Kicks, Rebounds, and Particle Production}},  {\em Phys. Rev.}
  {\bf D89} (2014), no.~8 084074, [\href{http://arxiv.org/abs/1310.5149}{{\tt
  arXiv:1310.5149}}].

\bibitem{Padilla:2015wlv}
A.~Padilla, E.~Platts, D.~Stefanyszyn, A.~Walters, A.~Weltman, and T.~Wilson,
  {\it {How to Avoid a Swift Kick in the Chameleons}},  {\em JCAP} {\bf 1603}
  (2016), no.~03 058, [\href{http://arxiv.org/abs/1511.0576}{{\tt
  arXiv:1511.0576}}].

\bibitem{Brax:2008hh}
P.~Brax, C.~van~de Bruck, A.-C. Davis, and D.~J. Shaw, {\it {f(R) Gravity and
  Chameleon Theories}},  {\em Phys. Rev.} {\bf D78} (2008) 104021,
  [\href{http://arxiv.org/abs/0806.3415}{{\tt arXiv:0806.3415}}].

\bibitem{Hu:2007nk}
W.~Hu and I.~Sawicki, {\it {Models of f(R) Cosmic Acceleration that Evade
  Solar-System Tests}},  {\em Phys. Rev.} {\bf D76} (2007) 064004,
  [\href{http://arxiv.org/abs/0705.1158}{{\tt arXiv:0705.1158}}].

\bibitem{Joyce:2014kja}
A.~Joyce, B.~Jain, J.~Khoury, and M.~Trodden, {\it {Beyond the Cosmological
  Standard Model}},  {\em Phys. Rept.} {\bf 568} (2015) 1--98,
  [\href{http://arxiv.org/abs/1407.0059}{{\tt arXiv:1407.0059}}].

\bibitem{Song:2006ej}
Y.-S. Song, W.~Hu, and I.~Sawicki, {\it {The Large Scale Structure of f(R)
  Gravity}},  {\em Phys. Rev.} {\bf D75} (2007) 044004,
  [\href{http://arxiv.org/abs/astro-ph/0610532}{{\tt astro-ph/0610532}}].

\bibitem{Nojiri:2006be}
S.~Nojiri and S.~D. Odintsov, {\it {Modified gravity and its reconstruction
  from the universe expansion history}},  {\em J. Phys. Conf. Ser.} {\bf 66}
  (2007) 012005, [\href{http://arxiv.org/abs/hep-th/0611071}{{\tt
  hep-th/0611071}}].

\bibitem{Nojiri:2006gh}
S.~Nojiri and S.~D. Odintsov, {\it {Modified f(R) gravity consistent with
  realistic cosmology: From matter dominated epoch to dark energy universe}},
  {\em Phys. Rev.} {\bf D74} (2006) 086005,
  [\href{http://arxiv.org/abs/hep-th/0608008}{{\tt hep-th/0608008}}].

\bibitem{Pogosian:2007sw}
L.~Pogosian and A.~Silvestri, {\it {The pattern of growth in viable f(R)
  cosmologies}},  {\em Phys. Rev.} {\bf D77} (2008) 023503,
  [\href{http://arxiv.org/abs/0709.0296}{{\tt arXiv:0709.0296}}]. [Erratum:
  Phys. Rev.D81,049901(2010)].

\bibitem{Nojiri:2010wj}
S.~Nojiri and S.~D. Odintsov, {\it {Unified cosmic history in modified gravity:
  from F(R) theory to Lorentz non-invariant models}},  {\em Phys. Rept.} {\bf
  505} (2011) 59--144, [\href{http://arxiv.org/abs/1011.0544}{{\tt
  arXiv:1011.0544}}].

\bibitem{Lombriser:2014dua}
L.~Lombriser, {\it {Constraining chameleon models with cosmology}},  {\em
  Annalen Phys.} {\bf 526} (2014) 259--282,
  [\href{http://arxiv.org/abs/1403.4268}{{\tt arXiv:1403.4268}}].

\bibitem{Burrage:2014daa}
C.~Burrage, E.~J. Copeland, and J.~Stevenson, {\it {Ellipticity Weakens
  Chameleon Screening}},  {\em Phys. Rev.} {\bf D91} (2015) 065030,
  [\href{http://arxiv.org/abs/1412.6373}{{\tt arXiv:1412.6373}}].

\bibitem{Cabre:2012tq}
A.~Cabre, V.~Vikram, G.-B. Zhao, B.~Jain, and K.~Koyama, {\it {Astrophysical
  Tests of Modified Gravity: A Screening Map of the Nearby Universe}},  {\em
  JCAP} {\bf 1207} (2012) 034, [\href{http://arxiv.org/abs/1204.6046}{{\tt
  arXiv:1204.6046}}].

\bibitem{Jain:2011ji}
B.~Jain and J.~VanderPlas, {\it {Tests of Modified Gravity with Dwarf
  Galaxies}},  {\em JCAP} {\bf 1110} (2011) 032,
  [\href{http://arxiv.org/abs/1106.0065}{{\tt arXiv:1106.0065}}].

\bibitem{Burrage:2014oza}
C.~Burrage, E.~J. Copeland, and E.~A. Hinds, {\it {Probing Dark Energy with
  Atom Interferometry}},  {\em JCAP} {\bf 1503} (2015), no.~03 042,
  [\href{http://arxiv.org/abs/1408.1409}{{\tt arXiv:1408.1409}}].

\bibitem{Brax:2013cfa}
P.~Brax, G.~Pignol, and D.~Roulier, {\it {Probing Strongly Coupled Chameleons
  with Slow Neutrons}},  {\em Phys. Rev.} {\bf D88} (2013) 083004,
  [\href{http://arxiv.org/abs/1306.6536}{{\tt arXiv:1306.6536}}].

\bibitem{Elder:2016yxm}
B.~Elder, J.~Khoury, P.~Haslinger, M.~Jaffe, H.~Müller, and P.~Hamilton, {\it
  {Chameleon Dark Energy and Atom Interferometry}},  {\em Phys. Rev.} {\bf D94}
  (2016), no.~4 044051, [\href{http://arxiv.org/abs/1603.0658}{{\tt
  arXiv:1603.0658}}].

\bibitem{Schlogel:2015uea}
S.~Schlögel, S.~Clesse, and A.~Füzfa, {\it {Probing Modified Gravity with
  Atom-Interferometry: a Numerical Approach}},  {\em Phys. Rev.} {\bf D93}
  (2016), no.~10 104036, [\href{http://arxiv.org/abs/1507.0308}{{\tt
  arXiv:1507.0308}}].

\bibitem{Rider:2016xaq}
A.~D. Rider, D.~C. Moore, C.~P. Blakemore, M.~Louis, M.~Lu, and G.~Gratta, {\it
  {Search for Screened Interactions Associated with Dark Energy Below the 100
  $\mathrm{\mu m}$ Length Scale}},  {\em Phys. Rev. Lett.} {\bf 117} (2016),
  no.~10 101101, [\href{http://arxiv.org/abs/1604.0490}{{\tt
  arXiv:1604.0490}}].

\bibitem{Brax:2013uh}
P.~Brax, A.-C. Davis, and J.~Sakstein, {\it {Pulsar Constraints on Screened
  Modified Gravity}},  {\em Class. Quant. Grav.} {\bf 31} (2014) 225001,
  [\href{http://arxiv.org/abs/1301.5587}{{\tt arXiv:1301.5587}}].

\bibitem{Santos:2016rdg}
M.~Vargas~dos Santos and D.~F. Mota, {\it {Extrasolar planets as a probe of
  modified gravity}},  \href{http://arxiv.org/abs/1603.0324}{{\tt
  arXiv:1603.0324}}.

\bibitem{Jain:2012tn}
B.~Jain, V.~Vikram, and J.~Sakstein, {\it {Astrophysical Tests of Modified
  Gravity: Constraints from Distance Indicators in the Nearby Universe}},  {\em
  Astrophys. J.} {\bf 779} (2013) 39,
  [\href{http://arxiv.org/abs/1204.6044}{{\tt arXiv:1204.6044}}].

\bibitem{Freedman:2010xv}
W.~L. Freedman and B.~F. Madore, {\it {The Hubble Constant}},  {\em Ann. Rev.
  Astron. Astrophys.} {\bf 48} (2010) 673--710,
  [\href{http://arxiv.org/abs/1004.1856}{{\tt arXiv:1004.1856}}].

\bibitem{Vikram:2014uza}
V.~Vikram, J.~Sakstein, C.~Davis, and A.~Neil, {\it {Astrophysical Tests of
  Modified Gravity: Stellar and Gaseous Rotation Curves in Dwarf Galaxies}},
  \href{http://arxiv.org/abs/1407.6044}{{\tt arXiv:1407.6044}}.

\bibitem{Schmidt:2010jr}
F.~Schmidt, {\it {Dynamical Masses in Modified Gravity}},  {\em Phys. Rev.}
  {\bf D81} (2010) 103002, [\href{http://arxiv.org/abs/1003.0409}{{\tt
  arXiv:1003.0409}}].

\bibitem{Terukina:2013eqa}
A.~Terukina, L.~Lombriser, K.~Yamamoto, D.~Bacon, K.~Koyama, and R.~C. Nichol,
  {\it {Testing chameleon gravity with the Coma cluster}},  {\em JCAP} {\bf
  1404} (2014) 013, [\href{http://arxiv.org/abs/1312.5083}{{\tt
  arXiv:1312.5083}}].

\bibitem{Wilcox:2015kna}
H.~Wilcox et~al., {\it {The XMM Cluster Survey: Testing chameleon gravity using
  the profiles of clusters}},  {\em Mon. Not. Roy. Astron. Soc.} {\bf 452}
  (2015), no.~2 1171--1183, [\href{http://arxiv.org/abs/1504.0393}{{\tt
  arXiv:1504.0393}}].

\bibitem{Adelberger:2006dh}
E.~G. Adelberger, B.~R. Heckel, S.~A. Hoedl, C.~D. Hoyle, D.~J. Kapner, and
  A.~Upadhye, {\it {Particle Physics Implications of a Recent Test of the
  Gravitational Inverse Sqaure Law}},  {\em Phys. Rev. Lett.} {\bf 98} (2007)
  131104, [\href{http://arxiv.org/abs/hep-ph/0611223}{{\tt hep-ph/0611223}}].

\bibitem{Mota:2006ed}
D.~F. Mota and D.~J. Shaw, {\it {Strongly coupled chameleon fields: New
  horizons in scalar field theory}},  {\em Phys. Rev. Lett.} {\bf 97} (2006)
  151102, [\href{http://arxiv.org/abs/hep-ph/0606204}{{\tt hep-ph/0606204}}].

\bibitem{Mota:2006fz}
D.~F. Mota and D.~J. Shaw, {\it {Evading Equivalence Principle Violations,
  Cosmological and other Experimental Constraints in Scalar Field Theories with
  a Strong Coupling to Matter}},  {\em Phys. Rev.} {\bf D75} (2007) 063501,
  [\href{http://arxiv.org/abs/hep-ph/0608078}{{\tt hep-ph/0608078}}].

\bibitem{Upadhye:2012fz}
A.~Upadhye, {\it {Particles and forces from chameleon dark energy}},  in {\em
  {8th Patras Workshop on Axions, WIMPs and WISPs (AXION-WIMP 2012) Chicago,
  Illinois, July 18-22, 2012}}, 2012.
\newblock \href{http://arxiv.org/abs/1211.7066}{{\tt arXiv:1211.7066}}.

\bibitem{Upadhye:2012qu}
A.~Upadhye, {\it {Dark energy fifth forces in torsion pendulum experiments}},
  {\em Phys. Rev.} {\bf D86} (2012) 102003,
  [\href{http://arxiv.org/abs/1209.0211}{{\tt arXiv:1209.0211}}].

\bibitem{Lamoreaux:1996wh}
S.~K. Lamoreaux, {\it {Demonstration of the Casimir force in the 0.6 to 6
  micrometers range}},  {\em Phys. Rev. Lett.} {\bf 78} (1997) 5--8. [Erratum:
  Phys. Rev. Lett.81,5475(1998)].

\bibitem{Lambrecht:2011qm}
A.~Lambrecht and S.~Reynaud, {\it {Casimir and short-range gravity tests}},  in
  {\em {Experimental Gravity and Gravitational Waves, p.199-206 (Th\'e Gioi,
  2011)}}, 2011.
\newblock \href{http://arxiv.org/abs/1106.3848}{{\tt arXiv:1106.3848}}.

\bibitem{Brax:2007vm}
P.~Brax, C.~van~de Bruck, A.-C. Davis, D.~F. Mota, and D.~J. Shaw, {\it
  {Detecting chameleons through Casimir force measurements}},  {\em Phys. Rev.}
  {\bf D76} (2007) 124034, [\href{http://arxiv.org/abs/0709.2075}{{\tt
  arXiv:0709.2075}}].

\bibitem{Brax:2014zta}
P.~Brax and A.-C. Davis, {\it {Casimir, Gravitational and Neutron Tests of Dark
  Energy}},  {\em Phys. Rev.} {\bf D91} (2015), no.~6 063503,
  [\href{http://arxiv.org/abs/1412.2080}{{\tt arXiv:1412.2080}}].

\bibitem{Decca:2007yb}
R.~S. Decca, D.~Lopez, E.~Fischbach, G.~L. Klimchitskaya, D.~E. Krause, and
  V.~M. Mostepanenko, {\it {Tests of new physics from precise measurements of
  the Casimir pressure between two gold-coated plates}},  {\em Phys. Rev.} {\bf
  D75} (2007) 077101, [\href{http://arxiv.org/abs/hep-ph/0703290}{{\tt
  hep-ph/0703290}}].

\bibitem{Lambrecht:2005km}
A.~Lambrecht, V.~V. Nesvizhevsky, R.~Onofrio, and S.~Reynaud, {\it {Development
  of a high-sensitivity torsional balance for the study of the Casimir force in
  the 1-10 micrometre range}},  {\em Class. Quant. Grav.} {\bf 22} (2005)
  5397--5406.

\bibitem{Lamoreaux:2005zza}
S.~K. Lamoreaux and W.~T. Buttler, {\it {Thermal noise limitations to force
  measurements with torsion pendulums: Applications to the measurement of the
  Casimir force and its thermal correction}},  {\em Phys. Rev.} {\bf E71}
  (2005) 036109.

\bibitem{Brax:2010xx}
P.~Brax, C.~van~de Bruck, A.~C. Davis, D.~J. Shaw, and D.~Iannuzzi, {\it
  {Tuning the Mass of Chameleon Fields in Casimir Force Experiments}},  {\em
  Phys. Rev. Lett.} {\bf 104} (2010) 241101,
  [\href{http://arxiv.org/abs/1003.1605}{{\tt arXiv:1003.1605}}].

\bibitem{Almasi:2015zpa}
A.~Almasi, P.~Brax, D.~Iannuzzi, and R.~I.~P. Sedmik, {\it {Force sensor for
  chameleon and Casimir force experiments with parallel-plate configuration}},
  {\em Phys. Rev.} {\bf D91} (2015), no.~10 102002,
  [\href{http://arxiv.org/abs/1505.0176}{{\tt arXiv:1505.0176}}].

\bibitem{Geraci:2010ft}
A.~A. Geraci, S.~B. Papp, and J.~Kitching, {\it {Short-range force detection
  using optically-cooled levitated microspheres}},  {\em Phys. Rev. Lett.} {\bf
  105} (2010) 101101, [\href{http://arxiv.org/abs/1006.0261}{{\tt
  arXiv:1006.0261}}].

\bibitem{Burrage:2015lya}
C.~Burrage and E.~J. Copeland, {\it {Using Atom Interferometry to Detect Dark
  Energy}},  {\em Contemp. Phys.} {\bf 57} (2016), no.~2 164--176,
  [\href{http://arxiv.org/abs/1507.0749}{{\tt arXiv:1507.0749}}].

\bibitem{Hamilton:2015zga}
P.~Hamilton, M.~Jaffe, P.~Haslinger, Q.~Simmons, H.~Müller, and J.~Khoury, {\it
  {Atom-interferometry constraints on dark energy}},  {\em Science} {\bf 349}
  (2015) 849--851, [\href{http://arxiv.org/abs/1502.0388}{{\tt
  arXiv:1502.0388}}].

\bibitem{Brax:2010gp}
P.~Brax and C.~Burrage, {\it {Atomic Precision Tests and Light Scalar
  Couplings}},  {\em Phys. Rev.} {\bf D83} (2011) 035020,
  [\href{http://arxiv.org/abs/1010.5108}{{\tt arXiv:1010.5108}}].

\bibitem{Jaeckel:2010xx}
J.~Jaeckel and S.~Roy, {\it {Spectroscopy as a test of Coulomb's law: A Probe
  of the hidden sector}},  {\em Phys. Rev.} {\bf D82} (2010) 125020,
  [\href{http://arxiv.org/abs/1008.3536}{{\tt arXiv:1008.3536}}].

\bibitem{Schwob:1999zz}
C.~Schwob, L.~Jozefowski, B.~de~Beauvoir, L.~Hilico, F.~Nez, L.~Julien,
  F.~Biraben, O.~Acef, J.~J. Zondy, and A.~Clairon, {\it {Optical Frequency
  Measurement of the S-2- D-12 Transitions in Hydrogen and Deuterium: Rydberg
  Constant and Lamb Shift Determinations}},  {\em Phys. Rev. Lett.} {\bf 82}
  (1999) 4960--4963.

\bibitem{Simon:1980hu}
G.~G. Simon, C.~Schmitt, F.~Borkowski, and V.~H. Walther, {\it {Absolute
  electron Proton Cross-Sections at Low Momentum Transfer Measured with a High
  Pressure Gas Target System}},  {\em Nucl. Phys.} {\bf A333} (1980) 381--391.

\bibitem{Brax:2011hb}
P.~Brax and G.~Pignol, {\it {Strongly Coupled Chameleons and the Neutronic
  Quantum Bouncer}},  {\em Phys. Rev. Lett.} {\bf 107} (2011) 111301,
  [\href{http://arxiv.org/abs/1105.3420}{{\tt arXiv:1105.3420}}].

\bibitem{Ivanov:2012cb}
A.~N. Ivanov, R.~Hollwieser, T.~Jenke, M.~Wellenzohen, and H.~Abele, {\it
  {Influence of the chameleon field potential on transition frequencies of
  gravitationally bound quantum states of ultracold neutrons}},  {\em Phys.
  Rev.} {\bf D87} (2013), no.~10 105013,
  [\href{http://arxiv.org/abs/1207.0419}{{\tt arXiv:1207.0419}}].

\bibitem{Nesvizhevsky:2002ef}
V.~V. Nesvizhevsky et~al., {\it {Quantum states of neutrons in the Earth's
  gravitational field}},  {\em Nature} {\bf 415} (2002) 297--299.

\bibitem{Jenke:2014yel}
T.~Jenke et~al., {\it {Gravity Resonance Spectroscopy Constrains Dark Energy
  and Dark Matter Scenarios}},  {\em Phys. Rev. Lett.} {\bf 112} (2014) 151105,
  [\href{http://arxiv.org/abs/1404.4099}{{\tt arXiv:1404.4099}}].

\bibitem{Brax:2014gja}
P.~Brax, {\it {Testing Chameleon Fields with Ultra Cold Neutron Bound States
  and Neutron Interferometry}},  {\em Phys. Procedia} {\bf 51} (2014) 73--77.

\bibitem{Lemmel:2015kwa}
H.~Lemmel, P.~Brax, A.~N. Ivanov, T.~Jenke, G.~Pignol, M.~Pitschmann,
  T.~Potocar, M.~Wellenzohn, M.~Zawisky, and H.~Abele, {\it {Neutron
  Interferometry constrains dark energy chameleon fields}},  {\em Phys. Lett.}
  {\bf B743} (2015) 310--314, [\href{http://arxiv.org/abs/1502.0602}{{\tt
  arXiv:1502.0602}}].

\bibitem{Li:2016tux}
K.~Li et~al., {\it {Neutron Limit on the Strongly-Coupled Chameleon Field}},
  {\em Phys. Rev.} {\bf D93} (2016), no.~6 062001,
  [\href{http://arxiv.org/abs/1601.0689}{{\tt arXiv:1601.0689}}].

\bibitem{Brax:2009ey}
P.~Brax, C.~Burrage, A.-C. Davis, D.~Seery, and A.~Weltman, {\it {Higgs
  production as a probe of Chameleon Dark Energy}},  {\em Phys. Rev.} {\bf D81}
  (2010) 103524, [\href{http://arxiv.org/abs/0911.1267}{{\tt
  arXiv:0911.1267}}].

\bibitem{Brax:2010uq}
P.~Brax, C.~Burrage, A.-C. Davis, D.~Seery, and A.~Weltman, {\it {Anomalous
  coupling of scalars to gauge fields}},  {\em Phys. Lett.} {\bf B699} (2011)
  5--9, [\href{http://arxiv.org/abs/1010.4536}{{\tt arXiv:1010.4536}}].

\bibitem{Steffen:2010ze}
{\bf GammeV} Collaboration, J.~H. Steffen, A.~Upadhye, A.~Baumbaugh, A.~S.
  Chou, P.~O. Mazur, R.~Tomlin, A.~Weltman, and W.~Wester, {\it {Laboratory
  constraints on chameleon dark energy and power-law fields}},  {\em Phys. Rev.
  Lett.} {\bf 105} (2010) 261803, [\href{http://arxiv.org/abs/1010.0988}{{\tt
  arXiv:1010.0988}}].

\bibitem{Rybka:2010ah}
{\bf ADMX} Collaboration, G.~Rybka et~al., {\it {A Search for Scalar Chameleons
  with ADMX}},  {\em Phys. Rev. Lett.} {\bf 105} (2010) 051801,
  [\href{http://arxiv.org/abs/1004.5160}{{\tt arXiv:1004.5160}}].

\bibitem{Anastassopoulos:2015yda}
{\bf CAST} Collaboration, V.~Anastassopoulos et~al., {\it {Search for
  chameleons with CAST}},  {\em Phys. Lett.} {\bf B749} (2015) 172--180,
  [\href{http://arxiv.org/abs/1503.0456}{{\tt arXiv:1503.0456}}].

\bibitem{Burrage:2008ii}
C.~Burrage, A.-C. Davis, and D.~J. Shaw, {\it {Detecting Chameleons: The
  Astronomical Polarization Produced by Chameleon-like Scalar Fields}},  {\em
  Phys. Rev.} {\bf D79} (2009) 044028,
  [\href{http://arxiv.org/abs/0809.1763}{{\tt arXiv:0809.1763}}].

\bibitem{Burrage:2009mj}
C.~Burrage, A.-C. Davis, and D.~J. Shaw, {\it {Active Galactic Nuclei Shed
  Light on Axion-like-Particles}},  {\em Phys. Rev. Lett.} {\bf 102} (2009)
  201101, [\href{http://arxiv.org/abs/0902.2320}{{\tt arXiv:0902.2320}}].

\bibitem{Pettinari:2010ay}
G.~W. Pettinari and R.~Crittenden, {\it {On the Evidence for Axion-like
  Particles from Active Galactic Nuclei}},  {\em Phys. Rev.} {\bf D82} (2010)
  083502, [\href{http://arxiv.org/abs/1007.0024}{{\tt arXiv:1007.0024}}].

\bibitem{Amendola:2012ys}
{\bf Euclid Theory Working Group} Collaboration, L.~Amendola et~al., {\it
  {Cosmology and fundamental physics with the Euclid satellite}},  {\em Living
  Rev. Rel.} {\bf 16} (2013) 6, [\href{http://arxiv.org/abs/1206.1225}{{\tt
  arXiv:1206.1225}}].

\bibitem{Brax:2012gr}
P.~Brax, A.-C. Davis, B.~Li, and H.~A. Winther, {\it {A Unified Description of
  Screened Modified Gravity}},  {\em Phys. Rev.} {\bf D86} (2012) 044015,
  [\href{http://arxiv.org/abs/1203.4812}{{\tt arXiv:1203.4812}}].

\bibitem{Brax:2013yja}
P.~Brax, A.-C. Davis, and J.~Sakstein, {\it {Dynamics of Supersymmetric
  Chameleons}},  {\em JCAP} {\bf 1310} (2013) 007,
  [\href{http://arxiv.org/abs/1302.3080}{{\tt arXiv:1302.3080}}].

\bibitem{Brax:2012mq}
P.~Brax, A.-C. Davis, and J.~Sakstein, {\it {SUPER-Screening}},  {\em Phys.
  Lett.} {\bf B719} (2013) 210--217,
  [\href{http://arxiv.org/abs/1212.4392}{{\tt arXiv:1212.4392}}].

\bibitem{Vikram:2013uba}
V.~Vikram, A.~Cabré, B.~Jain, and J.~T. VanderPlas, {\it {Astrophysical Tests
  of Modified Gravity: the Morphology and Kinematics of Dwarf Galaxies}},  {\em
  JCAP} {\bf 1308} (2013) 020, [\href{http://arxiv.org/abs/1303.0295}{{\tt
  arXiv:1303.0295}}].

\bibitem{wald2010general}


\bibitem{Sakstein:2014isa}
J.~Sakstein, {\it {Disformal Theories of Gravity: From the Solar System to
  Cosmology}},  {\em JCAP} {\bf 1412} (2014) 012,
  [\href{http://arxiv.org/abs/1409.1734}{{\tt arXiv:1409.1734}}].

\bibitem{Sakstein:2015jca}
J.~Sakstein and S.~Verner, {\it {Disformal Gravity Theories: A Jordan Frame
  Analysis}},  {\em Phys. Rev.} {\bf D92} (2015), no.~12 123005,
  [\href{http://arxiv.org/abs/1509.0567}{{\tt arXiv:1509.0567}}].

\bibitem{Einstein:1938yz}
A.~Einstein, L.~Infeld, and B.~Hoffmann, {\it {The Gravitational equations and
  the problem of motion}},  {\em Annals Math.} {\bf 39} (1938) 65--100.

\bibitem{Damour:1986ny}
T.~Damour, {\it {THE PROBLEM OF MOTION IN NEWTONIAN AND EINSTEINIAN GRAVITY}},
  in {\em {300 Years of Gravity: A Conference to Mark the 300th Anniversary of
  the Publication of Newton's Principia Cambridge, England, June 29-July 4,
  1987}}, 1986.

\bibitem{Padmanabhan:2010zzb}
T.~Padmanabhan, {\em {Gravitation: Foundations and frontiers}}.
\newblock 2010.

\end{thebibliography}\endgroup

\end{document}